\documentclass{jkas}

\def\year{2026} % publication year
\def\volume{---} % journal volume
\def\issue{---} % journal issue
\def\beginpage{1} % first page of article
\def\received{November 14, 2025} % date paper was received by JKAS
\def\accepted{February 9, 2026} % date of acceptance
\def\published{---} % date of publication
\date{Received \received; Accepted \accepted; Published \published}

\newcommand\ion[2]{{#1}\,{\sc #2}} % ions: \ion{C}{iv} = C IV

\graphicspath{{./}{figures/}}

\title{%
K-DRIFT Science Theme: Galaxies in the Faint Universe
}

\author[1$\star$]{Woowon Byun}{0000-0002-7762-7712}
\author[1,2$\star$]{Yongmin Yoon}{0000-0003-0134-8968}
\author[1,3]{Jongwan Ko}{0000-0002-9434-5936}
\author[2]{Yun Hee Lee}{0000-0003-2779-6793}
\author[4]{Gain Lee}{0009-0007-8443-3143}
\author[4,5]{Ho Seong Hwang}{0000-0003-3428-7612}
\author[6]{Cristiano G. Sabiu}{0000-0002-5513-5303}
\author[1,3]{Kwang-il Seon}{0000-0001-9561-8134}
\author[1]{Kyungwon Chun}{0000-0001-9544-7021}
\author[1,3]{Jihye Shin}{0000-0001-5135-1693}
\author[1,7]{Jinsu Rhee}{0000-0002-0184-9589}
\author[1]{Jae-Woo Kim}{0000-0002-1710-4442}
\author[1,8]{Jaewon Yoo}{0000-0002-6841-8329}
\author[1]{Jaehyun Lee}{0000-0002-6810-1778}
\author[1]{Sang-Hyun Chun}{0000-0002-6154-7558}
\author[1,3]{Hong Soo Park}{0000-0002-3505-3036}
\author[1]{Soung-Chul Yang}{0000-0001-9842-639X}
\author[1]{Sungryong Hong}{0000-0001-9991-8222}
\author[1,3]{Jeehye Shin}{0000-0002-9914-3129}
\author[9]{Hyowon Kim}{0000-0003-4032-8572}

\affil[1]{Korea Astronomy and Space Science Institute, Daejeon 34055, Republic of Korea}
\affil[2]{Department of Astronomy and Atmospheric Sciences, Kyungpook National University, Daegu 41566, Republic of Korea}
\affil[3]{Department of Astronomy and Space Science, University of Science and Technology, Daejeon 34113, Republic of Korea}
\affil[4]{Astronomy Program, Department of Physics and Astronomy, Seoul National University, Seoul 08826, Republic of Korea}
\affil[5]{SNU Astronomy Research Center, Seoul National University, Seoul 08826, Republic of Korea}
\affil[6]{Natural Science Research Institute, University of Seoul, Seoul 02504, Republic of Korea}
\affil[7]{Department of Astronomy and Yonsei University Observatory, Yonsei University, Seoul 03722, Republic of Korea}
\affil[8]{Quantum Universe Center, Korea Institute for Advanced Study, Seoul 02455, Republic of Korea}
\affil[9]{Departamento de Física, Universidad Técnica Federico Santa María, Avenida España 1680, Valparaíso, Chile}
\def\corrauthor{%
W. Byun, \email{wbyun87@gmail.com}; Y. Yoon, \email{yyoon@knu.ac.kr}
}

\def\runningauthor{%
Byun et al.
}

\def\runningtitle{%
K-DRIFT Science Theme: Galaxies in the Faint Universe
}

\def\keywords{%
galaxies: general --- galaxies: interactions --- galaxies: stellar content --- galaxies: structure --- galaxies: dwarf galaxies --- galaxies: low surface brightness galaxies
}

\def\abstracttext{%
Low-surface-brightness (LSB) structures serve as evidence of the intricate mass assembly of galaxies, and dedicatedly studying them promises to give us profound insights into the evolutionary history of galaxies. Furthermore, delving into the properties of star formation (SF) in the LSB regime can broaden our understanding of SF activity in regions characterized by low surface gas density, thereby shedding light on fundamental cosmic processes. However, systematic uncertainties may hamper the exploration of the LSB universe by limiting detectable SB levels. Indeed, despite dedicated advancements in telescope and observing techniques over decades, achieving ultra-deep photometric depths in optical wavelengths remains a formidable challenge. To overcome this challenge and explore the LSB universe that we have yet to see, we have been developing a novel telescope called K-DRIFT. This paper outlines the telescope's specification and describes various LSB features we aim for, explicitly focusing on nearby individual galaxies. To further advance the capabilities of the K-DRIFT survey, focused on LSB detection, we present several feasible research topics that utilize other survey data together and discuss the role of LSB observation in understanding the evolution of galaxies. 
}

\begin{document}
\jkashead %% set title, authors, abstract, etc.

%%%%%%%%%%%%%%%%%%%%%%%%%%%%%%%%%%%%%%%%%%%%%%%%%%%%%%%%%%%%%%%%%%%%%
%%% BEGIN MAIN TEXT HERE %%%%%%%%%%%%%%%%%%%%%%%%%%%%%%%%%%%%%%%%%%%%
%%%%%%%%%%%%%%%%%%%%%%%%%%%%%%%%%%%%%%%%%%%%%%%%%%%%%%%%%%%%%%%%%%%%%

\section{Introduction} \label{sec:intro}
In a hierarchical galaxy formation framework, most galaxies undergo at least one interaction with other galaxies during their lifetime \cite[][]{1978MNRAS.183..341W,1991ApJ...379...52W}. This process, such as merger and/or accretion, may leave various tidal features \cite[][]{1988ApJ...331..699B,1992ApJ...399L.117H,2007ApJ...666...20P,2008ApJ...684.1062F,2008ApJ...689..936J,2013MNRAS.434.3348C,2014MNRAS.444..237P} that are typically in a low-surface-brightness (LSB) regime, i.e., $\mu_V \gtrsim 27$ mag arcsec$^{-2}$. Due to their long-lasting shape-retaining properties over several Gyr and their causalities, they can serve as evidence of the mass assembly history of galaxies \cite[][]{2001ApJ...548...33B,2008MNRAS.391...14D,2008ApJ...689..936J}. In addition to tidal features, various LSB objects exist around galaxies, and they can serve to verify the $\Lambda$ cold dark matter ($\Lambda$CDM) paradigm. For instance, comparing cosmological simulation results with a census of faint, low-mass satellite galaxies can help reinforce the standard cosmological model \cite[e.g.,][]{1999ApJ...522...82K,2007ApJ...670..313S,2013ApJ...765...22B,2018NatAs...2..162P,2022ApJ...936...38N}. Hence, LSB imaging has emerged as a crucial technique offering unique insights and a deeper understanding of the formation and evolution of galaxies. 

One way to hunt LSB features is by counting resolved stars directly in the vicinity of nearby galaxies with ground-based telescopes \cite[e.g.,][]{2007ApJ...658..337B,2014ApJ...787...19M,2018ApJ...868...55M,2023ApJ...949L..37S}
and the Hubble Space Telescope \cite[HST; e.g.,][]{2011ApJS..195...18R,2017MNRAS.466.1491H}. This method enables us to detect them more intuitively and investigate the details of the stellar population. However, it has the disadvantage of not being able to provide larger samples for statistical studies because it cannot be applied to relatively distant galaxies. 

Alternatively, one can use the integrated light of the LSB features to further explore the universe. The Sloan Digital Sky Survey (SDSS), one of the well-known conventional imaging surveys, has been limited in its exploration of the LSB universe due to its shallow photometric depth ($\mu_g\lesssim25$ mag arcsec$^{-2}$). This risks providing limited information from the brightest LSB structures, which result from the most massive progenitors \cite[see][]{2022MNRAS.513.1459M}, potentially leading to a biased and incomplete understanding of galaxy evolution. For this reason, there have recently been intensive attempts to achieve deeper SB limit and observe the LSB features of galaxies through an optimized observation strategy and data reduction \cite[e.g.,][]{2013ApJ...762...82M,2014ApJ...782L..24V,2016ApJ...830...62M,2016ApJ...833..168M,2016ApJ...823..123T,2016ApJ...826...59W,2018AJ....156..249B,2020ApJ...891...18B,2022PASP..134h4101B,2018A&A...614A.143M,2023A&A...669A.103M}. 

However, even these dedicated imaging surveys can pose significant challenges in the extremely LSB domain ($\mu_r\gtrsim30$ mag arcsec$^{-2}$). Detecting such faint features is highly sensitive to even minor uncertainties, making the careful reduction of systematic uncertainties crucial. Two primary factors contribute to systematic uncertainty: (1) the varying night sky brightness affected by moonlight, airglow, zodiacal light, and light pollution, and (2) instrumental artifacts such as light scattering and reflection. These factors not only degrade data quality but also interfere with accurate data processing. 

The Dragonfly Telephoto Array \cite[][]{2014PASP..126...55A} is one of the specialized LSB imaging systems. It consists of multiple telephoto lenses, which enhance light-gathering capability and reduce instrumental artifacts by simplifying optics design, thereby improving sensitivity to LSB detection. This system facilitated unprecedented photometric depth and the discovery of many LSB features in the nearby universe \cite[e.g.,][]{2014ApJ...787L..37M,2016ApJ...830...62M,2016ApJ...833..168M,2014ApJ...782L..24V,2015ApJ...798L..45V,2019ApJ...883L..32V}. 

Similarly, the Korea Astronomy and Space Science Institute (KASI) has launched an ambitious project named ``KASI Deep Rolling Imaging Fast Telescope (K-DRIFT),'' which involves developing an innovative telescope and conducting an LSB imaging survey. A pilot study using the prototype telescope has demonstrated the feasibility of manufacturing such a telescope and confirmed that efficient LSB observation is achievable \cite[see][]{2022PASP..134h4101B}. These findings confirm that the extensive LSB survey we are pursuing is promising. 

We are currently developing an upgraded K-DRIFT telescope, with plans to conduct an LSB imaging survey targeting the southern sky. This paper is part of a series that outlines the scientific objectives we aim to achieve and discusses the implications related to the LSB features of nearby individual galaxies. The paper is organized as follows: Section \ref{sec:telescope} describes the telescope specifications and observation strategy. In Section \ref{sec:targets}, we present the LSB features that can be discovered and studied. Section \ref{sec:impact} explores the synergies between K-DRIFT survey data and other multi-wavelength data in the context of galaxy evolution research.

%% The "t!" tells LaTeX to put the figure "here" first, at the "top" next
%% and to override the normal way of calculating a float position
\begin{figure}[t!]
\includegraphics[width=\linewidth]{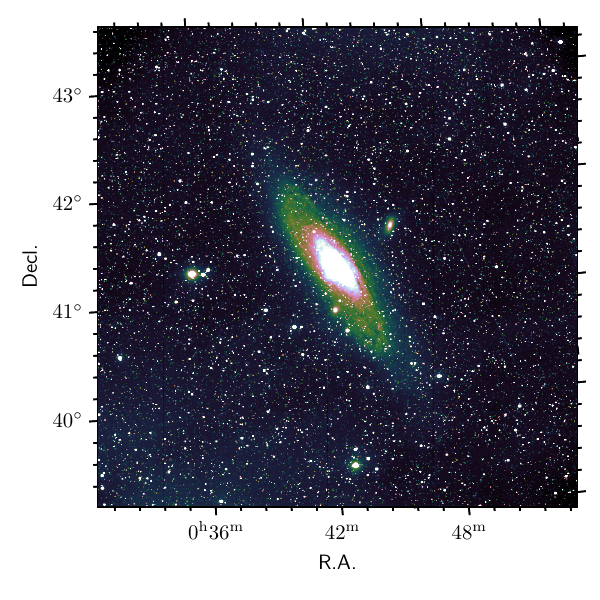}
\centering
\caption{Example observation of the Andromeda Galaxy (M31) obtained with K-DRIFT G1. This single image was captured using the $L$ filter with an exposure time of 300 sec. No data reduction or post-processing was applied, but a false-color (cubehelix) scheme was used for visual clarity.
\label{fig:f0}}
\end{figure} 

\section{Telescope Specifications and Observation Strategy} \label{sec:telescope}
The K-DRIFT survey is designed as a deep, wide-field mission to provide a systematic census of LSB features across the nearby universe. The primary strategy is to cover a contiguous footprint within the declination range of $-40^\circ$ to $-20^\circ$, achieving a 3$\sigma$ SB limit of $\sim$29--30 mag arcsec$^{-2}$ in $10^{\prime\prime}\times10^{\prime\prime}$ scales. This survey establishes a foundational dataset for detecting LSB objects, providing the broad observational context for specific scientific programs, such as the study of nearby massive galaxy samples described in Section \ref{sec:targets}. 

To achieve these goals, the K-DRIFT telescope adopts an off-axis freeform three-mirror system designed to be free of linear astigmatism \cite[][]{10.1117/12.2023433,2020PASP..132d4504P}. The performance of this optics system was assessed through on-sky test observations using a prototype telescope named K-DRIFT Pathfinder, which was found to be suitable for LSB observations despite its small aperture size \cite[see][]{2022PASP..134h4101B}. To significantly scale up the survey throughput and further suppress scattered light beyond the prototype phase, we have developed an advanced version named K-DRIFT Generation 1 (G1). 

K-DRIFT G1 consists of two identical telescopes. Each telescope has a 500-mm physical aperture, but an effective entrance pupil diameter of 300 mm. With a focal length of 1050 mm, this corresponds to an effective focal ratio of $f$/3.5. Each telescope is equipped with a sCMOS camera (Teledyne COSMOS-66) composed of a $\mathrm{8k}\times\mathrm{8k}$ single chip with a pixel scale of $\sim$2$^{\prime\prime}$. Due to a wide field of view (FoV) of $\sim$$4.5^\circ\times4.5^\circ$ (Figure \ref{fig:f0}), it enables us to explore the sky fast and efficiently. Three optical broadband ($ugr$) filters and one Luminance ($L$) filter are available for each telescope to discover LSB objects and infer their physical properties from their colors. After several years of imaging surveys, we also plan to utilize narrowband filters, such as H$\alpha$ and/or [\ion{O}{iii}], to investigate recently star-forming or ionized gas regions. Detailed specifications of the telescope and survey strategy can be found in \cite{2025arXiv251022250K}. 

\begin{figure}[t!]
\includegraphics[width=\linewidth]{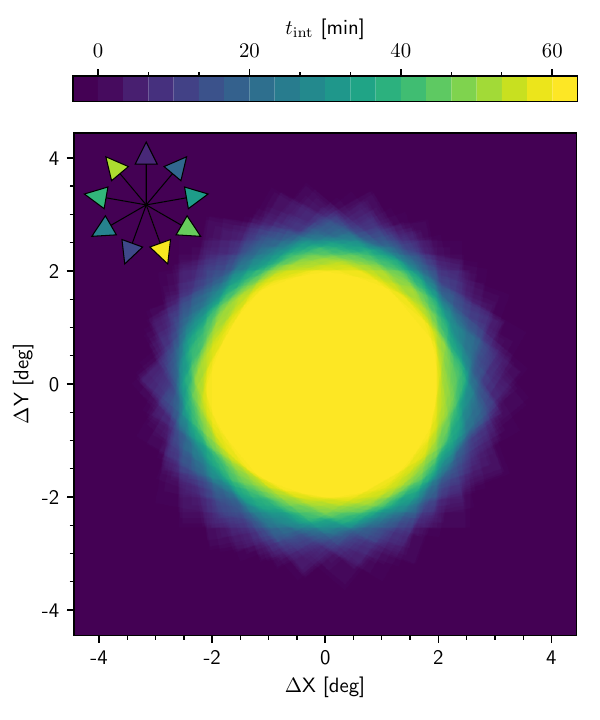}
\centering
\caption{Coverage of a single observation sequence using the rolling dithering method. The color represents the integration time. Since the dithering offset is up to $\sim$40$^\prime$, the effective area with the longest integration time is $\sim$11.5 deg$^2$. The camera's position angle rotating by 160$^\circ$ is also shown in the upper left corner. 
\label{fig:f1}}
\end{figure}

The observations will be carried out using the ``rolling dithering'' method for targets of interest. This technique involves repeatedly capturing multiple exposures along a random-offset dithering pattern while also rotating the camera's position angle between exposures. Figure \ref{fig:f1} illustrates the resulting coverage of observations using the rolling dithering method. Although this observation strategy may be inefficient for covering wide areas since the stacked final image is circular, it has the advantage of providing a uniform coverage of the effective area of $\sim$11.5 deg$^2$ centered on the targets. 

%% The "t!" tells LaTeX to put the figure "here" first, at the "top" next
%% and to override the normal way of calculating a float position

%% The "t!" tells LaTeX to put the figure "here" first, at the "top" next
%% and to override the normal way of calculating a float position
\begin{figure*}[t!]
\includegraphics[width=\linewidth]{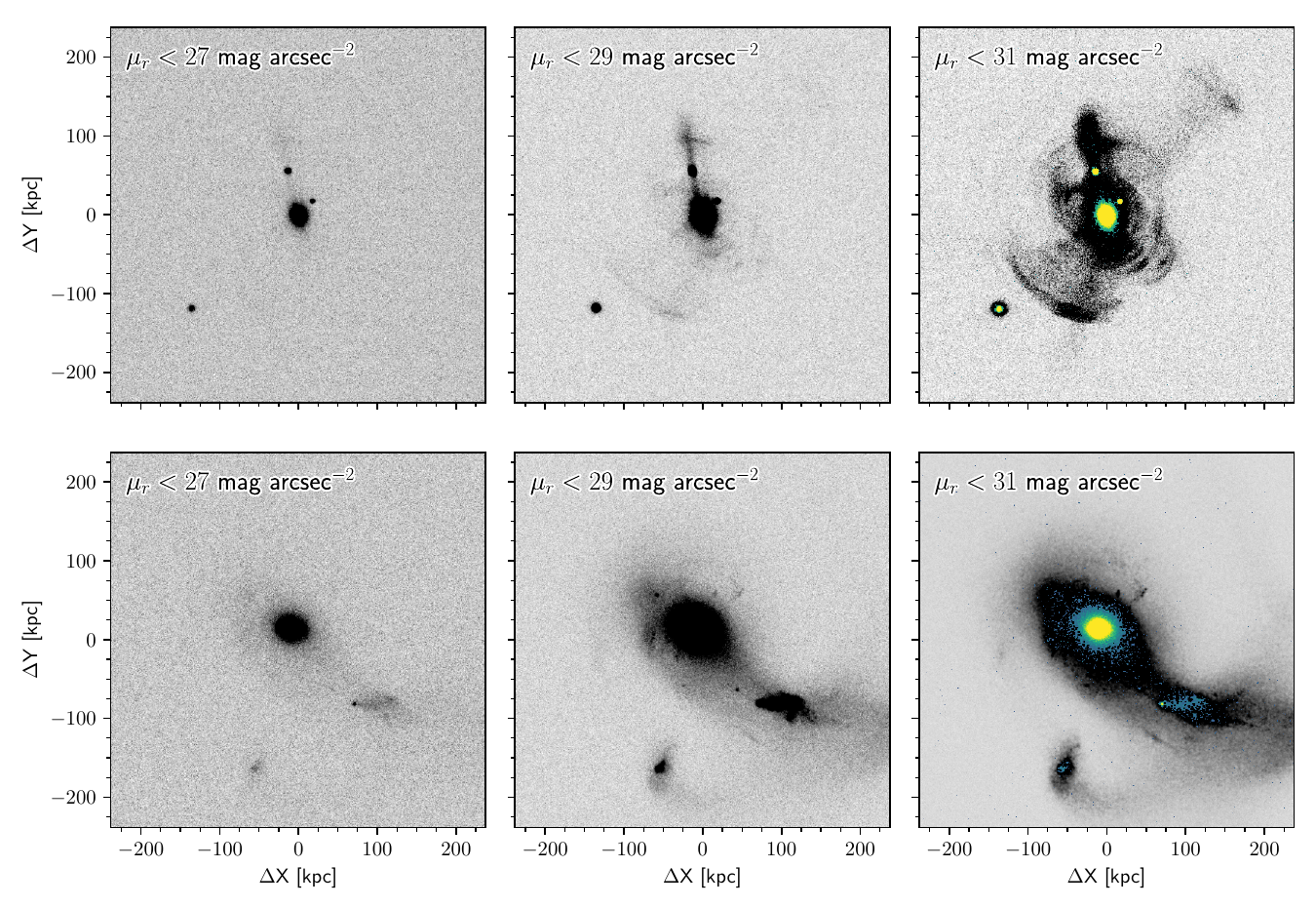}
\centering
\caption{Stellar structures of merging galaxies adopted from the GRT (top) and NewHorizon2 (bottom) simulations. The SB limits of each panel are denoted in the upper left corner. The color-coded pixels in the right panel represent the pixels above 27 mag arcsec$^{-2}$, which is the same as in the left panel.
\label{fig:f2}}
\end{figure*}

The basic procedure for a single observation sequence is as follows: 
\begin{enumerate}
\item{Fix the camera's position angle and conduct multiple exposures along an offset dithering pattern. For each exposure, the offset is randomly determined within a $40^\prime\times40^\prime$ box\footnote{This is as large as approximately eight times the size of Milky Way (MW)-like galaxies at 20 Mpc.} relative to the reference coordinate.}
\item{Rotate the camera's position angle by 160$^\circ$ relative to the previous angle.}
\item{Repeat the above two steps.}
\end{enumerate}
By default, a single observation sequence comprises nine rotations and a total integration time of $\sim$1 hr. That is, when the exposure time per frame is 40 sec, the camera rotates after every 10 exposures. Multiple visits to the target field may be required to achieve the desired photometric depth. 

Note that the rolling dithering method not only ensures maximum on-source integration time but also enables more accurate data processing \cite[see][]{2016ApJ...823..123T}. For instance, a master flat frame for flat-fielding is usually generated by combining twilight or night sky images. However, the resulting master flat may not be perfect since the sky contains non-negligible fluctuations and gradients. The rolling dithering method can average out these fluctuations when creating a master flat frame, thereby reducing flat-fielding errors \cite[see details in][]{2025PASP..137e4502B}. 

\section{Primary Targets of Scientific Research} \label{sec:targets}
LSB observations are a crucial approach for inferring the hierarchical evolution of galaxies, as LSB features associated with a host galaxy provide decisive evidence of its mass assembly history. Figure \ref{fig:f2} shows two representative examples of merging galaxies drawn from different cosmological simulations. The upper panels are adopted from the Galaxy Refinement Technique (GRT) simulation\footnote{This simulation is an optimized tool for exploring the formation and evolution of LSB features in galaxy clusters and components therein.} \cite[][]{2022ApJ...925..103C}. In this example, the host galaxy with a stellar mass of $\sim$$1.3\times10^{11}$ $M_\odot$ has experienced two mergers: a 1:4 major merger 11 Gyr ago and a 1:10 minor merger 5 Gyr ago. The lower panels are adopted from the NewHorizon2 simulation (\citealt{2024ApJS..271....1Y}; Yi et al. in preparation). In this case, the host galaxy with a stellar mass of $\sim$$5.6\times10^{10}$ $M_\odot$ has experienced two major mergers 5 and 3 Gyr ago. Together, these examples demonstrate that extended and diffuse LSB features spanning several hundred kpc are a common outcome of hierarchical mass assembly, and that detecting and correctly interpreting such structures require observations reaching sufficiently deep SB limits, at least $\sim$29--30 mag arcsec$^{-2}$ at 3$\sigma$ level. 

K-DRIFT will focus primarily on detecting and classifying LSB features that have yet to be discovered, using $gr$- and $L$-filter imaging data. By leveraging the diverse morphologies of merger remnants and their varying survival times, we can infer their origins and reconstruct the merger histories of their host galaxies. In addition, we can identify numerous dwarf galaxies that either result from the merger process or may contribute to the formation of future LSB structures around host galaxies. This section outlines the key targets we will investigate and provides a brief description of their characteristics (Section \ref{sec:targets:tidal}--\ref{sec:targets:outskirts}). Furthermore, given the anticipated growth of the dataset, we outline the potential role of advanced methods, such as machine learning (ML), as part of the overall framework for the efficient and systematic identification of LSB objects (Section \ref{sec:targets:ml}). 

To address these scientific objectives within the observational framework established in Section \ref{sec:telescope}, the K-DRIFT survey is anchored by a well-defined sample of nearby galaxies. Within the survey footprint ($-40^\circ < \textrm{Decl.} < -20^\circ$), we prioritize $\sim$400 nearby galaxies as primary targets, drawn from existing catalogs such as the Carnegie-Irvine Galaxy Survey \cite[][]{2011ApJS..197...21H} and curated void-galaxy samples \cite[][]{2019MNRAS.482.4329P}. This target selection spans a wide range of galaxy masses and environments, enabling systematic investigations of LSB features. While these prioritized galaxies serve as the main scientific focus, we also anticipate serendipitous discoveries of LSB objects distributed throughout the field beyond the primary target sample owing to the wide FoV. 

Note that the incidence and detectability of individual LSB features involve a complex interplay of galaxy properties and observational conditions, resulting in a wide range of reported occurrence rates in the literature that depend on data quality and methodology \cite[see][]{2019A&A...632A.122M,2022MNRAS.513.1459M,2023ApJ...951..137K,2025arXiv250922802C}. While the deeper imaging of K-DRIFT is expected to substantially enhance detectability relative to existing large-area surveys such as SDSS or the DESI Legacy Imaging Surveys,\footnote{\url{https://legacysurvey.org}} these coupled dependencies make deterministic yield predictions inherently uncertain. 

Nevertheless, we provide quantitative references based on the literature at the end of several subsequent subsections to provide a baseline for comparison. However, these values must be interpreted with extreme caution; they are intended to serve only as contextual references or conservative lower and upper bounds for scale, based on current simulations and observations, rather than definitive predictions for the K-DRIFT survey. Ultimately, K-DRIFT aims to empirically resolve these uncertainties by establishing more robust constraints on the quantitative estimation of LSB features in the nearby universe. 

\subsection{Tidal Structures as Merger Remnants} \label{sec:targets:tidal}
Galaxies undergoing merger events leave distinct LSB features in their surroundings, such as stellar streams and tidal tails \cite[][]{2001ApJ...548...33B,2005ApJ...635..931B,2010AJ....140..962M,2023A&A...671A.141M}. These structures may originate from different sources, such as progenitors, mechanisms, and epochs; therefore, classifying them will enable us to investigate the evolution of galaxies effectively. 

Although many researchers strive to develop automated approaches for classifying tidal structures \cite[e.g.,][]{2019A&A...626A..49P,2019MNRAS.483.2968W,2020ApJ...895..115F,2022MNRAS.513.1459M,2024MNRAS.534.1459G}, numerous studies still depend heavily on visual inspection \cite[e.g.,][]{2015MNRAS.446..120D,2022A&A...662A.124S,2023MNRAS.520.1155J,2023A&A...671A.141M,2024MNRAS.532..883S,2024A&A...686A.182V,2024ApJ...974..299Y}. However, visual inspection often introduces subjectivity, potentially leading to inconsistencies among researchers. This issue may also arise from the inherent limitation of observations that rely on snapshots taken at single epochs and from a single viewing angle. 

Therefore, this section describes the selection criteria for identifying and classifying prominent tidal features (e.g., stellar streams, tidal tails, and shells) as well as less prominent ones. Note that the overall content of this section is mainly based on \cite{2019A&A...632A.122M}, \cite{2022A&A...662A.124S}, and \cite{2023MNRAS.520.5870G}. 

\subsubsection{Stellar Streams} \label{sec:targets:tidal:streams}
Stellar streams typically refer to thin and elongated tidal features. Since they are likely to be associated with minor mergers \cite[][]{2005ApJ...635..931B,2006ApJ...642L.137B,2010AJ....140..962M}, they appear distinct from the host galaxies, which are only slightly disturbed. \cite{2019A&A...632A.122M} showed that the lifetime of stellar streams depends on the incidence of merger events but is typically $\sim$3 Gyr. They also noted that stellar streams are the most frequent features in merging systems, although these structures are often difficult to detect observationally due to their faintness. 

\begin{figure}[t!]
\includegraphics[width=\linewidth]{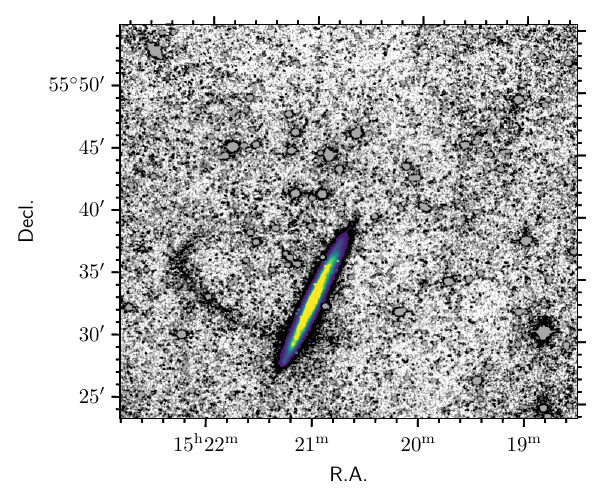}
\centering
\caption{Stellar stream structure associated with NGC 5907 \cite[][]{2022PASP..134h4101B}. Bright sources are masked, while color-coded pixels highlight the disk of NGC 5907 brighter than 27 mag arcsec$^{-2}$. A prominent, curved stellar stream extends to the east of the galaxy. 
\label{fig:f3}}
\end{figure}

Figure \ref{fig:f3} presents the observation result for the stellar stream of NGC 5907 obtained with K-DRIFT Pathfinder \cite[see][]{2022PASP..134h4101B}. As the detectable SB limit reaches approximately 28 mag arcsec$^{-2}$, the stellar stream to the east of the galaxy becomes visible. Note that \cite{2019ApJ...883L..32V} demonstrated that this structure extends across the galaxy toward the northwest. 

%% The "t!" tells LaTeX to put the figure "here" first, at the "top" next
%% and to override the normal way of calculating a float position

Since the K-DRIFT survey aims to reach a photometric depth of $\sim$30 mag arcsec$^{-2}$, we expect to detect such giant structures extending over hundreds of kpc around galaxies within $\sim$100 Mpc volume. Since the progenitors of stellar streams are likely to be less massive and to have relatively simple stellar populations, deep observations using the $gr$- and $L$-filters will allow us to infer their properties, including at least the total stellar mass. Furthermore, as recently demonstrated by \cite{2023ApJ...954..195N} and \cite{2025ApJ...994...36W}, the morphology (or orbit) of stellar streams offers an avenue to investigate the shape of DM halos of their host galaxies. 

\subsubsection{Tidal Tails} \label{sec:targets:tidal:tails}
Unlike stellar streams, tidal tails can be characterized by elongated stellar structures that are likely to originate from the host galaxies. In other words, stellar streams are associated with the accretion of small galaxies, while tidal tails are considered ejected stellar components from host galaxies. The formation of tidal tails is generally attributed to intermediate-mass and major merger events, during which the secondary galaxy can disrupt the host galaxy. Since tidal tails are formed during intense interaction phases, they tend to be relatively bright and easy to detect. While minor mergers might also produce tidal tails, these tend to be smaller and fainter, making detection more challenging \cite[][]{2008ApJ...684.1062F}. 

Simulation studies have shown that at least two tails form in the post-merger stage, with a typical survival time of $\sim$2 Gyr \cite[e.g.,][]{1972ApJ...178..623T,1995AJ....110..140H,2019A&A...632A.122M}. The shape of tidal tails can vary depending on the internal kinematics of the involved galaxies, whether they are rotation- or dispersion-dominated. For example, mergers in disk-disk galaxies leave long and curved structures, while mergers in pairs of early-type galaxies (ETGs) tend to generate broader ones, i.e., fan-like structures \cite[][]{2008MNRAS.388.1537M,2009AJ....138.1417T,2023MNRAS.520.5870G}. 

\begin{figure}[t!]
\includegraphics[width=\linewidth]{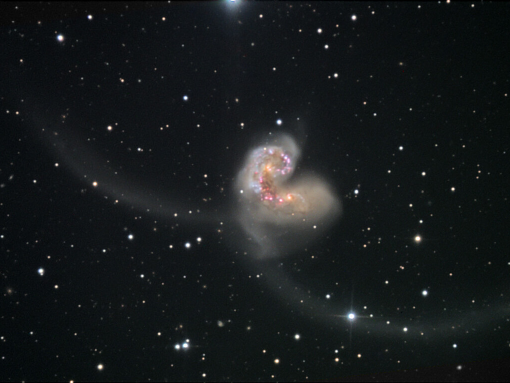}
\centering
\caption{Tidal tail structures within the Antennae Galaxies, which are in the process of colliding with each other. The FoV of the image is $\sim$$16^\prime\times12^\prime$. Credit: KPNO/NOIRLab/NSF/AURA/Bob and Bill Twardy/Adam Block
\label{fig:f4}}
\end{figure}

Figure \ref{fig:f4} shows the major merging system of the two spiral galaxies NGC 4038 and NGC 4039, also known as the Antennae Galaxies. This example highlights the importance of deep, wide-field imaging to capture tidal tails that extend several times beyond their host galaxies. The K-DRIFT telescope is an observation system that meets this requirement. Within $\sim$100 Mpc volume, we expect to observe a number of galaxies with diverse tidal tails. If a merger occurred long ago, the tidal tails would likely appear significantly fainter. Detecting such faint features can provide valuable clues about relatively old merger histories. 

%% The "t!" tells LaTeX to put the figure "here" first, at the "top" next
%% and to override the normal way of calculating a float position

\subsubsection{Shells} \label{sec:targets:tidal:shells}
Shells are arc-shaped features that are typically concentric with the host galaxy. They can be formed during both minor and major mergers, and in particular, are deeply related to radial encounters \cite[][]{1986A&A...166...53D,1988ApJ...331..682H,1989ApJ...342....1H,1990dig..book...72P,2012ApJ...753...43K,2013arXiv1312.1643E,2015MNRAS.446..120D,2017Galax...5...34P,2018MNRAS.480.1715P,2019MNRAS.487..318K,2023MNRAS.520.5870G}. In most cases, multiple shells appear across a wide range of galactocentric radii, as stellar components disrupted during multiple encounters accumulate at specific radii \cite[][]{1983ApJ...274..534M,2019A&A...632A.122M,2023MNRAS.520.5870G}. \cite{2019A&A...632A.122M} presented that their lifetime is derived to be $\sim$4 Gyr from a single merger event and may be shortened if additional violent mergers occur. Detection of shell structures is less sensitive to the SB limit because of their sharp-edged features, but it is highly dependent on the projection effect. 

\begin{figure}[t!]
\includegraphics[width=\linewidth]{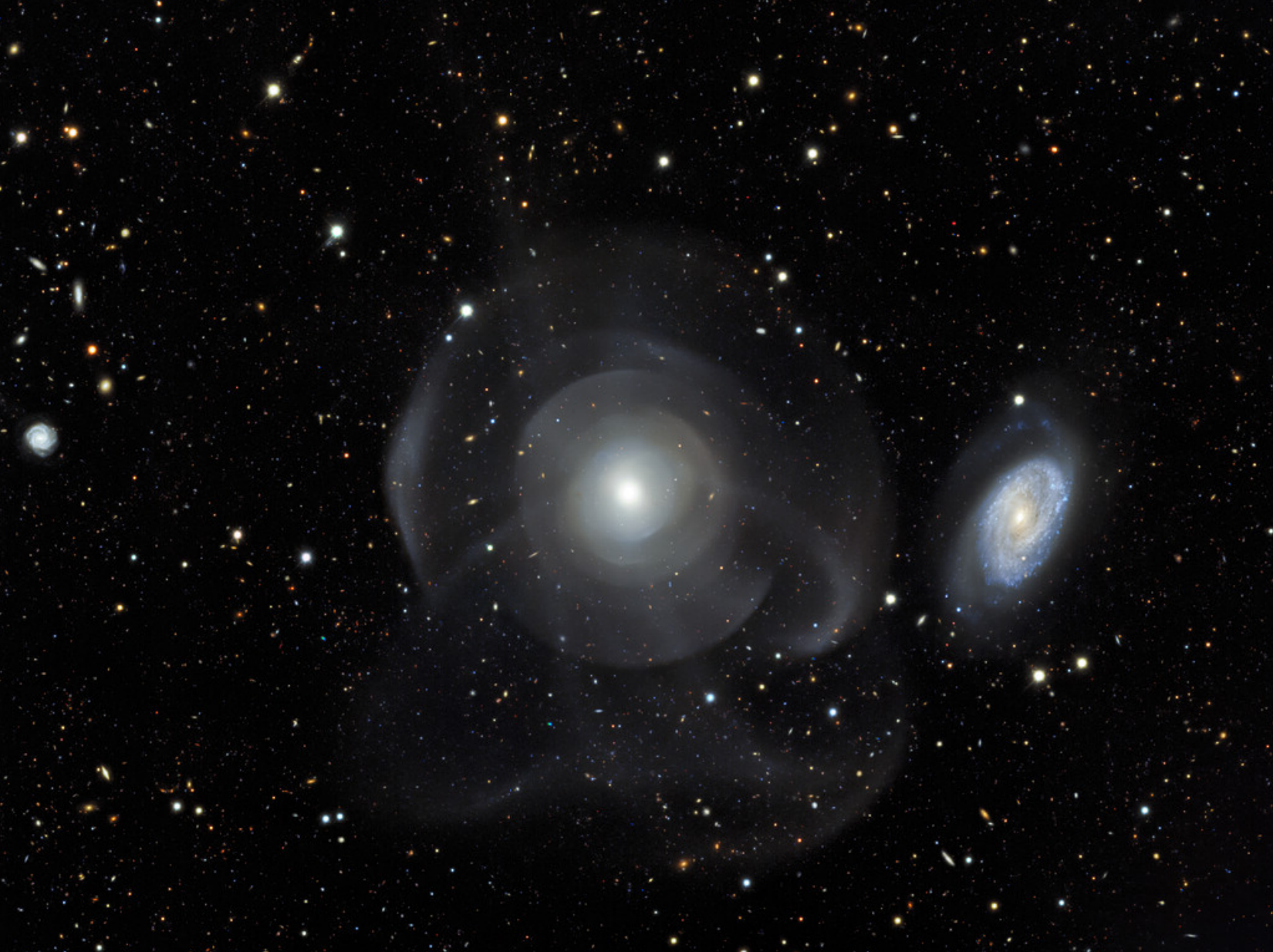}
\centering
\caption{Shell structures within NGC 474, which is considered to have experienced recent mergers or interactions with smaller galaxies. The FoV of the image is $\sim$$18^\prime\times13^\prime$. Credit: DES/DOE/Fermilab/NCSA \& CTIO/NOIRLab/NSF/AURA 
\label{fig:f5}}
\end{figure}

Figure \ref{fig:f5} shows the shell structures in NGC 474. The shells are the most noticeable features, and other tidal features are also visible. Although classifying galaxy morphology is often challenging due to the mixing of such structures \cite[e.g.,][]{1991rc3..book.....D,1994AJ....108.2128C,2011AJ....141...74H}, shells are known to be particularly prominent in ETGs at a given photometric depth \cite[][]{2023MNRAS.520.5870G,2024ApJ...965..158Y}. In other words, merger events that create shell structures may be closely linked to the morphological transformation of galaxies into ETGs. This connection allows us to study the properties of the progenitors of shell structures and explore the evolutionary processes driving morphological changes, such as examining the relationship between galaxy morphology and the prevalence of shell structures. 

%% The "t!" tells LaTeX to put the figure "here" first, at the "top" next
%% and to override the normal way of calculating a float position

\subsubsection{Other Less Prominent Features} \label{sec:targets:tidal:others}
Although confidence in detection and classification may be relatively low, several internal structures within galaxies could suggest traces of galaxy interaction. For example, numerous galaxies have been found to exhibit X-shaped structures in their central region \cite[e.g.,][]{1995ApJ...447L..87M,2009AJ....138.1417T}. These structures, which can technically be formed via the buckling instability of the bar \cite[][]{1981A&A....96..164C,1991Natur.352..411R,2015MNRAS.448..713P}, may also be induced by minor mergers or perturbations caused by companion galaxies \cite[][]{1988ApJ...324..741W,1995ApJ...447L..87M}. Another is the disk-like structures in the central region of ETGs, which are proposed to originate from gas-rich mergers \cite[][]{2008MNRAS.391.1137L}. In addition, twisted or warped isophotes may serve as evidence of galaxy interaction \cite[][]{1983MNRAS.203P..19G,2011MNRAS.414.2923K,2015MNRAS.454..299S,2018ApJ...856...11G}. The lopsidedness of stellar components and hydrogen gas in the galaxy outskirts can also be used as a tracer of their early assembly history \cite[][]{2023MNRAS.526..567D}.

Since these features are embedded within the galaxy's light, additional techniques, such as galaxy model subtraction, are required to identify them reliably \cite[see][]{2023MNRAS.520.5870G}. Given the limited spatial resolution of the K-DRIFT telescope, detecting these features in abundance would be more challenging than detecting the three prominent structures mentioned above. Nevertheless, exploring the presence of any companion galaxies involved in forming those features could provide valuable insights. 

\paragraph{Literature-based quantitative references:} \cite{2024MNRAS.530.4422K} examined mock images---mimicking the Legacy Survey of Space and Time \cite[][]{2019ApJ...873..111I}---generated from several simulations and reported that $\sim$30--40\% of galaxies with stellar mass $9.5\le \mathrm{log}\ (M_\star/M_\odot)\le12.6$ exhibit tidal features, including prominent structures and asymmetric outskirts. More specifically, $\sim$11\% have stellar streams and/or tidal tails, $\sim$0.5\% have shells, $\sim$30\% exhibit asymmetric outskirts, and $\sim$17\% have double cores. Observationally, \cite{2024ApJ...974..299Y} focused on massive ETGs and reported that nearly half of their sample shows detectable tidal features. 

\subsection{Stellar Halos} \label{sec:targets:halo}
The distinct tidal structures described in Section \ref{sec:targets:tidal} will be disrupted and eventually form well-mixed stellar halos surrounding their host galaxies \cite[][]{2008MNRAS.391...14D,2008ApJ...689..936J,2017MNRAS.466.1491H,2018MNRAS.474.5300D,2018A&A...614A.143M,2019MNRAS.485.2589M}. Since these stellar halos are relics of ancient mass assembly events, their characteristics, such as color, shape, and mass fraction, provide important insights into the evolutionary history of individual galaxies. 

Interestingly, several studies have reported substantial variations in both the mass fraction and the shape of stellar halos in nearby galaxies \cite[e.g.,][]{2016ApJ...830...62M,2019MNRAS.490.1539R}. Some galaxies have even exhibited a lack of stellar halo components, which appears to differ from the predictions of $\Lambda$CDM models \cite[e.g.,][]{2016ApJ...830...62M,2022MNRAS.513.1459M}. From this, one may infer that galaxies undergo a wide variety of evolutionary processes. In contrast, \cite{2019MNRAS.487.1580W} argued that the scatter in stellar halo profiles across galaxies disappears when the profiles are scaled by the virial radii of their DM halos. 

Such debate may be attributed to insufficient observational data, either quantitatively or qualitatively, given the diffuse and faint nature of stellar halos. To address these challenges, the following factors should be taken into consideration. First, a deeper photometric depth must be achieved than that required to detect other tidal features. Detecting stellar halos typically requires an SB level of at least $\sim$27--28 mag arcsec$^{-2}$ \cite[][]{2011ApJ...739...20C,2012arXiv1204.3082B}, with reliable confirmation only possible at SB limits 5--10 times fainter \cite[][]{2013MNRAS.434.3348C,2016ApJ...830...62M,2021A&A...654A..40T}. Second, uncertainties that can arise during data processing must be minimized as much as possible. Specifically, the accuracy of sky subtraction and background determination can significantly affect the analysis of stellar halos. Furthermore, accurately estimating the properties of stellar halos requires a FoV substantially larger than their angular extent. At this stage, capturing the extended structures of stellar halos along with the sky background in a single frame could be more advantageous than mosaicking. 

The K-DRIFT survey is well-suited for investigating stellar halos because it satisfies the abovementioned conditions, i.e., deep photometric depth and a wide FoV. We expect that systematic exploration with deep and homogeneous imaging data will enable the identification of stellar halos and the investigation of their properties in many galaxies within a $\sim$100 Mpc volume. 

It is worth noting that an extended point-spread-function (PSF) wing can affect the measurement of the shape of stellar halos and the amount of their light \cite[see][]{2008MNRAS.388.1521D,2013MNRAS.431.1121T,2014A&A...567A..97S,2015A&A...577A.106S,2016ApJ...823..123T,2019MNRAS.487.1580W}. A recent evaluation of K-DRIFT Pathfinder revealed that the PSF wing is very compact; its intensity decreases by approximately six orders of magnitude at a 1$^\prime$ radius compared to the central region. We predict that K-DRIFT G1 will achieve comparable or better results. While the impact of the PSF wing varies depending on the angular sizes and central brightness of target galaxies, regions fainter than 28 mag arcsec$^{-2}$ are likely to be affected. Post-processing techniques, such as PSF deconvolution, are also planned to address this concern. 

\paragraph{Literature-based quantitative references:} \cite{2016ApJ...830...62M} showed that MW-like galaxies exhibit substantial diversity in stellar halo mass fraction, with values as low as $\sim$0.01\%, whereas earlier simulation studies predicted stellar halo mass fractions of $\sim$1\% \cite[e.g.,][]{2010MNRAS.406..744C,2013MNRAS.434.3348C,2015ApJ...799..184P}. Because this result is based on a relatively small sample of 10 galaxies, including the MW and M31, there remains room for more systematic studies with larger samples to better understand the nature of stellar halos. 

\subsection{Dwarf Galaxies} \label{sec:targets:dwf}
Dwarf galaxies, characterized by relatively small size and low luminosity, play a crucial role in our understanding of galaxy formation and evolution, serving as building blocks of larger galaxies in hierarchical structure formation \cite[see][]{1980lssu.book.....P}. The abundance and spatial distribution of dwarf galaxies allow us to verify the $\Lambda$CDM models \cite[see][]{1999ApJ...522...82K,2005Natur.435..629S}. In addition, owing to their low metallicity and limited dust content, dwarf galaxies are ideal targets for investigating stellar properties and evolution \cite[see][]{1998ARA&A..36..435M,2009ARA&A..47..371T}. Examination of dwarf galaxies in diverse environments can shed light on the environmental effect in galaxy evolution since they are susceptible to the surrounding environment \cite[see][]{1996MNRAS.278..947B,2015A&A...573A..48P}. 

One of our main goals is to discover faint dwarf galaxies in diverse environments, thereby enhancing observational completeness. We will also conduct in-depth studies on the origins of dwarf galaxies with peculiar properties. This section outlines several key targets of our research. 

\subsubsection{Satellite Dwarf Galaxies} \label{sec:targets:dwf:sat}
In the $\Lambda$CDM paradigm, the large-scale universe is well predicted, while the much smaller-scale features are often poorly reproduced \cite[][]{2017ARA&A..55..343B}. The discrepancy between observations and theory on small-scale structures is also called the ``dwarf galaxy problems,'' with some examples of this tension provided below:
\begin{itemize}
\item{Missing satellites problem: the $\Lambda$CDM models predict the existence of a large number of low-mass DM halos that are expected to host dwarf galaxies. However, the observed number of satellite dwarf galaxies around massive galaxies appears significantly fewer than predicted \cite[][]{,1999ApJ...522...82K,1999ApJ...524L..19M}.}
\item{Too-big-to-fail problem: many of the satellite galaxies in the $\Lambda$CDM models are predicted to reside in massive subhalos, implying that they should host numerous stars and thus be easily observable. However, these bright satellite dwarf galaxies have not been detected \cite[][]{2011MNRAS.415L..40B,2014MNRAS.444..222G}.}
\item{Plane of satellites problem: satellite galaxies are often distributed in a noticeably flattened shape. This plane-like alignment of satellite galaxies contradicts the prediction of the $\Lambda$CDM models, which expect them to be more randomly distributed around the host galaxy \cite[][]{2012MNRAS.423.1109P,2017A&A...602A.119M}.}
\end{itemize}

Much of our understanding of dwarf galaxy issues comes from observations of the MW and the Local Group, leaving open the possibility that these findings may simply reflect statistical anomalies. Although numerous studies have conducted extensive surveys for satellite dwarf galaxies \cite[e.g.,][]{2007ApJ...656L..13I,2013MNRAS.428..573S,2015AstBu..70..379K,2017ApJ...848...19P,2019ApJ...885...88P,2018ApJ...868...96C,2018ApJ...856...69D,2018ApJ...863..152S,2018ApJ...865..125T,2019A&A...629A..18M,2020ApJ...891...18B,2020ApJ...891..144C,2020A&A...644A..91M,2021MNRAS.507.4764G,2021A&A...654A..40T,2022ApJ...933...47C,2022ApJ...936...38N}, the sample sizes remain insufficient to reach a definitive consensus. 

One way to address these issues is by considering the possibility of observational incompleteness as demonstrated in numerical simulations \cite[e.g.,][]{2023ApJ...951..137K}. In this context, several scenarios have been proposed: the number of observable satellite dwarf galaxies may decrease if supernovae-driven outflows suppress their star formation (SF) activity \cite[e.g.,][]{2011MNRAS.417.1260F} or if tidal effects from nearby massive galaxies contribute to their disruption \cite[e.g.,][]{2004ApJ...609..482K,2017MNRAS.471.1709G}. 

Deep imaging surveys have the potential to significantly increase the number of previously undetected satellite dwarf galaxies. In addition, by detecting LSB remnants produced by tidal effects, we can directly investigate why satellite dwarf galaxies may appear less frequently. We anticipate that the K-DRIFT survey will discover numerous satellite dwarf galaxies around nearby massive host galaxies within $\sim$50 Mpc volume. This systematic exploration will help understand the nature of satellite dwarf galaxies and alleviate cosmological tensions. 

Finally, we introduce interesting research topics that can be attempted using the survey data. Several studies have suggested that the number of satellite galaxies is closely related to the properties of the host galaxies. For example, \cite{2005AJ....129..178K} reported a correlation between the abundance of satellite dwarf galaxies and the bulge properties of the host galaxies \cite[see also][]{2023A&A...678A..92M}. In addition, \cite{2022ApJ...930...69S} reported that there is a tight correlation between the number of satellite galaxies and the stellar mass of the dominant merger to the host galaxy, where the latter is defined as either the total stellar mass accreted to date (i.e., stellar halo mass) or the stellar mass of the most massive satellite expected to dominate the stellar halo in the future \cite[see also][]{2015MNRAS.448L..77D,2018MNRAS.474.5300D,2018NatAs...2..737D,2019MNRAS.485.2589M,2020ApJ...905...60S}. These results indicate that satellite galaxies can be used as proxies for estimating the properties of host galaxies, including merging history and/or DM halo mass. 

\paragraph{Literature-based quantitative references:} \cite{2021MNRAS.507.4211E} examined the TNG50 simulation data \cite[][]{2019ComAC...6....2N} and found that MW-like hosts contain a median of $5^{+6}_{-3}$ satellite galaxies with stellar masses $M_\star\ge8\times10^6 M_\odot$. Observationally, \cite{2024ApJ...976..117M} reported that 101 MW-mass systems host, on average, $\sim$3--4 satellite galaxies, with substantial scatter from zero to 13. While satellite detectability is affected by multiple factors beyond the SB limit, deeper imaging may improve completeness at the faint end. 

\subsubsection{Ultra-Diffuse Galaxies} \label{sec:targets:dwf:udg}
\begin{figure}[t!]
\includegraphics[width=\linewidth]{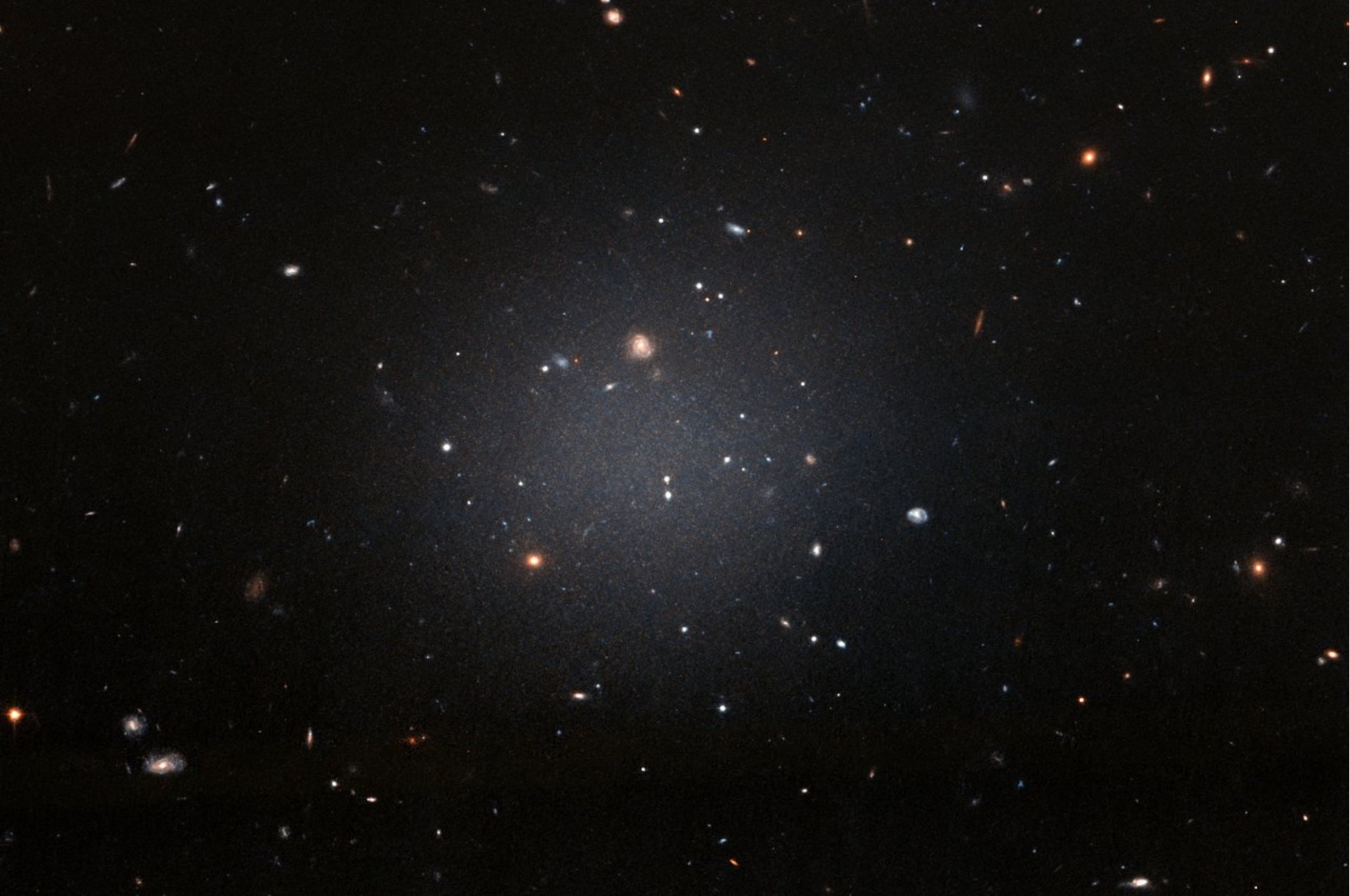}
\centering
\caption{NGC 1052-DF2: a UDG, also classified as a DMDG. The FoV of the image is $\sim$$2.5^\prime\times1.7^\prime$. Credit: NASA, ESA, and P. van Dokkum (Yale University)
\label{fig:f6}}
\end{figure}

Since the initial reports of faint galaxies, numerous LSB galaxies have been discovered over the decades \cite[e.g.,][]{1984AJ.....89..919S,1988ApJ...330..634I,1991ApJ...376..404B,1997AJ....114..635D,1997ARA&A..35..267I}. Recently, one of the most interesting discoveries is the identification of many very faint and extended objects in the Coma cluster \cite[][]{2015ApJ...798L..45V}. These objects were proposed to be a new class of giant spheroidal galaxies, known as ultra-diffuse galaxies (UDGs). As shown in Figure 
\ref{fig:f6}, they are comparable in size to the MW ($r_\mathrm{eff} \gtrsim 1.5$ kpc) but exhibit an extremely LSB ($\mu_{0,g} \gtrsim 24$ mag arcsec$^{-2}$) nature. 

%% The "t!" tells LaTeX to put the figure "here" first, at the "top" next
%% and to override the normal way of calculating a float position

With the rapid growth of the UDG samples in diverse environments from clusters to fields \cite[e.g.,][]{2015ApJ...807L...2K,2015ApJ...809L..21M,2016A&A...590A..20V,2016ApJS..225...11Y,2017ApJ...842..133L,2017A&A...608A.142V,2018ApJ...857..104G,2019ApJS..240....1Z,2023ApJS..267...27Z,2020ApJS..247...46B,2020A&A...642A..48I,2020ApJ...899...69L,2021A&A...654A.105M,2022MNRAS.510..946G,2023ApJS..267...10B}, an important question arises: ``How can they still survive in dense environments despite their appearance of being vulnerable to tidal interactions?'' This question has led to a debate about the origin of their formation. Two primary mechanisms for UDG formation have been proposed: puffed-up dwarfs caused by tidal interactions or stellar feedback \cite[e.g.,][]{2017MNRAS.466L...1D,2018MNRAS.480L.106O,2018ApJ...856L..31T,2018ApJ...866L..11B,2019MNRAS.485..382C,2021MNRAS.502..398C,2019MNRAS.487.5272J,2020MNRAS.494.1848S,2021ApJ...919...72J,2021MNRAS.503..679M,2023MNRAS.518.2497Z} versus failed MW-like galaxies that somehow could not form a stellar component sufficiently \cite[e.g.,][]{2016ApJ...822L..31P,2016ApJ...828L...6V,2018ApJ...856L..31T}. 

In this context, the total mass of UDGs is a fundamental factor in discriminating their possible formation mechanisms. One common method for estimating total mass relies on the abundance of globular clusters within these galaxies \cite[e.g.,][]{2017ApJ...844L..11V,2018ApJ...862...82L,2022MNRAS.510..946G,2022MNRAS.511.4633S}. More recently, radial mass segregation of globular clusters has also been employed to limit the mass and distribution of their DM halos hosting UDGs \cite[e.g.,][]{2024ApJ...964...53L}. 

Meanwhile, the formation mechanisms of UDGs may differ based on their environments. In dense regions, environmental processes, such as ram-pressure stripping or tidal heating, are often suggested. Conversely, in less-dense regions, intrinsic factors have also been proposed---for example, diffuse morphologies resulting from early, short starburst events \cite[][]{2021NatAs...5.1308G,2022ApJ...927L..28D} or peculiar physical characteristics, like the high spin of DM halos \cite[][]{2016MNRAS.459L..51A,2017MNRAS.470.4231R}. These ideas support the claim that UDGs can be part of the faint-end dwarf population. As a result, the definition of UDGs has even been broadened to include sub-UDGs \cite[$r_\mathrm{eff}\sim1.0$--1.5 kpc;][]{2023ApJS..267...10B} or reclassified as ultra-puffy galaxies, meaning outliers in the mass-size relation \cite[][]{2023ApJ...955....1L}. 

Indeed, the properties of UDGs in different environments are distinguishable. Compared to those in denser environments, UDGs in less-dense environments are more likely to have bluer colors \cite[][]{2016AJ....151...96M,2017MNRAS.468.4039R}, and some even exhibit emission lines \cite[][]{2017ApJ...838L..21K}. This variation suggests that UDGs may have mutually exclusive origins depending on their surroundings.  

In this regard, the K-DRIFT survey, which aims to target UDGs across a broad range of environments, will help clarify their origins. We anticipate discovering many UDG candidates within $\sim$50–100 Mpc volume. Based on this catalog, we will confirm their distances and total masses through follow-up spectroscopic observations. Additionally, \ion{H}{i} survey data may assist in studying the origins of UDGs. In the future, we plan to conduct intensive observations of \ion{H}{i}-bearing UDGs using $u$- or H$\alpha$-narrowband filters to explore the signs of recent SF activity. 

\paragraph{Literature-based quantitative references:} \cite{2018A&A...614A..21J} estimated a UDG number density of approximately $10^{-3}$ Mpc$^{-3}$ using data from the Arecibo Legacy Fast ALFA survey \cite[][]{2005AJ....130.2598G} and SDSS. Additionally, \cite{2023ApJS..265...57P} reported 2427, 2103, and 864 dwarf elliptical galaxies in two clusters, 265 groups, and 586 field environments, respectively. Of these, 12, 59, and 29 were classified as UDGs, suggesting that UDGs may constitute a few percent of the dwarf galaxy population. Moreover, our exploratory estimates based on TNG50 simulations suggest that up to several thousand diffuse objects may exist within the K-DRIFT survey footprint, providing an upper bound rather than a detection prediction. 

\subsubsection{Compact Elliptical Galaxies} \label{sec:targets:dwf:ce}
Compact elliptical galaxies (cEs) are a subclass of elliptical galaxies characterized by their small sizes (tens to hundreds pc), low stellar masses ($\sim$10$^{8-10}$ $M_\odot$), and high surface brightness \cite[][]{1973ApJ...179..423F,2014MNRAS.443.1151N}. They are considered as a population distinct from ordinary dwarf ETGs and have characteristics between the low-mass end of massive ETGs and ultra-compact dwarfs \cite[][]{2005ApJ...627..203H,2005A&A...430L..25M,2007A&A...466L..21C,2009MNRAS.397.1816P,2013MNRAS.430.1956H,2015ApJ...804...70G}. The origin of this rare population, with its ambiguous nature, is still debated \cite[see][]{2001ApJ...557L..39B,2019ApJ...875...58D,2020ApJ...903...65K}. 

One scenario proposes that cEs are formed by the tidal stripping of massive progenitors since they tend to have higher metallicity than other galaxies with similar stellar mass and to be located near massive host galaxies \cite[][]{2001ApJ...557L..39B,2002AJ....124..310C,2014MNRAS.443.1151N,2015Sci...348..418C,2016MNRAS.456..617J}. In addition, some cEs have been found to have disturbed outer disks or tidal streams, which can be direct evidence of this scenario \cite[][]{2002ApJ...568L..13G,2011MNRAS.414.3557H,2013ApJ...767..133P,2018MNRAS.473.1819F}. 

An alternative scenario proposes that cEs are actually classical elliptical galaxies with lower luminosity, as they follow the same trend on the fundamental plane as normal ellipticals \cite[][]{1984ApJ...282...85W,2009ApJ...695..101D,2012ApJS..198....2K,2015A&A...578A.134S}. In addition, the discovery of isolated cEs can also be a crucial clue that they are not formed only by environmental effects but can be born as intrinsically compact galaxies, e.g., through early mergers \cite[][]{2013MNRAS.430.1956H,2014MNRAS.443..446P,2021ApJ...917L...9R}. 

Not all cEs exhibit tidal features and/or have massive companion galaxies, suggesting that they may be formed via both channels \cite[see][]{2022ApJ...934L..35C,2023MNRAS.tmp.2247D}. However, we cannot rule out the possibility that tidal features have not been detected simply because of the limited photometric depth of the existing data. Through the K-DRIFT survey, we aim to examine the presence or absence of these LSB features, thereby providing strong constraints on their formation mechanisms.

\subsubsection{Dark-Matter-Deficient Galaxies} \label{sec:targets:dwf:dmdg}
In the $\Lambda$CDM paradigm, it is believed that DM constitutes the majority of the mass in the universe, and galaxies reside in DM halos. Recently, \cite{2018Natur.555..629V} reported the remarkable discovery that NGC 1052-DF2 is DM-deficient (see Figure \ref{fig:f6}), and soon after, they found that NGC 1052-DF4 also lacks DM \cite[][]{2019ApJ...874L...5V}.\footnote{This interpretation is accepted when their distance is $\sim$20 Mpc. The measurement of their distance remains disputed \cite[see][]{2019MNRAS.486.1192T,2021ApJ...914L..12S}.} These rare objects, dubbed DM-deficient galaxies (DMDGs), challenge our current understanding of hierarchical galaxy formation and evolution \cite[see][]{2019MNRAS.489.2634H}. Furthermore, the two galaxies have also been found to have unusual globular cluster properties \cite[see][]{2018ApJ...856L..30V,2021ApJ...909..179S}. This implies that the role of DM in galaxy formation may be more complex than previously believed. 

Meanwhile, DMDGs are thought to form through at least two mechanisms. One is tidal interactions by close encounters with massive galaxies \cite[][]{2019MNRAS.488.3298J,2021MNRAS.502.1785J,2022NatAs...6..496M,2022MNRAS.510.2724O}. \cite{2022NatAs...6..496M} predicted that one-third of massive central galaxies may host at least one satellite DMDG formed in this way. In this context, DMDGs can also be referred to as ``tidal dwarf galaxies'' because they are a type of tidal remnant. 

The other mechanism is in-situ SF in DM-free gas clouds \cite[][]{2020NatAs...4..246G,2022Natur.605..435V}, which does not necessarily require the condition of proximity to massive companions. Although the gas clouds can form in various ways, \cite{2022Natur.605..435V} recently proposed that high-velocity collisions of gas-rich progenitor galaxies could separate gas from DM halos, leading to the formation of low-luminosity galaxies in the resulting isolated gas clouds \cite[see also][]{2020ApJ...899...25S,2021ApJ...917L..15L}. Similarly, \cite{2023MNRAS.525.2535O} demonstrated that head-on collisions between gas-containing DM subhalos within a host halo frequently occur, and can lead to DMDG formation. 

Hunting more of these galaxies is essential to advancing our understanding of DMDG formation mechanisms. Observational findings, such as a tendency to reside near massive companions or align linearly along collision trajectories, could provide conclusive evidence. The K-DRIFT survey is expected to discover numerous faint galaxies, either directly through satellite galaxy searches or serendipitously. Follow-up spectroscopic observations can confirm whether they are DMDGs. At this stage, K-DRIFT's wide FoV will be invaluable; for instance, if some galaxies were aligned along potential collision trajectories, they would be easier to recognize. Such galaxies could be considered DMDG candidates and prioritized for spectroscopic observations. 

\paragraph{Literature-based quantitative references:} \cite{2019A&A...626A..47H} predicted a number density of $\sim$$2.3\times10^{-4}\ h^3$ cMpc$^{-3}$ for DMDG candidates, particularly systems with tidal dwarf galaxy formation channels, using the Illustris simulation.\footnote{\url{https://www.illustris-project.org}} Regarding satellite dwarf galaxies, \cite{2021MNRAS.502.1785J} reported that $\sim$10\% of dwarf satellite galaxies around massive hosts may exhibit a high degree of DM deficiency due to tidal stripping. In addition to these channels, DMDGs formed through high-velocity gas collisions \cite[see][]{2020ApJ...899...25S,2022Natur.605..435V} may exist, although such systems are expected to be rare. 

\subsubsection{Dark Galaxies} \label{sec:targets:dwf:dg}
Dark galaxies, the opposite of DMDGs, are hypothetical objects believed to consist almost entirely of DM with little or no stars. According to the galaxy formation scenario within the $\Lambda$CDM framework \cite[][]{1978MNRAS.183..341W}, dark galaxies are thought to be leftovers of DM halos that somehow failed to form stars. 

Dark galaxies can provide a unique laboratory for testing the $\Lambda$CDM models, thereby contributing to a deeper understanding of galaxy formation and evolution. They may also be helpful for understanding the ``missing satellites problem,'' mentioned in Section \ref{sec:targets:dwf:sat}. Ultimately, detecting dark galaxies would provide crucial evidence for the existence of DM and serve as a first step toward understanding its nature. 

Beginning with limited theoretical studies on the nature of dark galaxies \cite[e.g.,][]{1997MNRAS.292L...5J,2001MNRAS.322..658T}, recent numerical studies using cosmological hydrodynamical simulations have been conducted to investigate their origin and evolution \cite[e.g.,][]{2017MNRAS.465.3913B,2023MNRAS.519.1425P,2024ApJ...962..129L}. Observational studies, on the other hand, have faced challenges because dark galaxies are too faint in optical images. Since they are expected to have abundant gas reservoirs rather than stars, \ion{H}{i} surveys can serve as an alternative. Indeed, several dark galaxy candidates have been proposed based on \ion{H}{i} detections \cite[e.g.,][]{2005ApJ...622L..21M,2016ApJ...828L...6V,2020A&A...642L..10B,2021AJ....162..274L, 2023ApJ...944L..40X,2025ApJS..279...38K}. 

However, \ion{H}{i} detection alone is not enough to confirm the existence of dark galaxies, as such detections can be explained by intergalactic clouds produced by ram pressure stripping \cite[][]{2005A&A...437L..19O, 2021A&A...650A..99J} or tidal debris from galaxy-galaxy interactions \cite[][]{2005MNRAS.363L..21B,2008ApJ...673..787D,2010ApJ...717L.143M, 2022AJ....164..233T}. Even the presence of some bright stellar components does not guarantee their existence. For example, the candidate CDG-1 in the Perseus cluster was found to contain four GCs \cite[][]{2022ApJ...935....3L}, but the most definitive feature---the diffuse light component---has not yet been clearly identified \cite[][]{2024RNAAS...8..135V}, leaving its classification as a dark galaxy unconfirmed. 

Consequently, deep optical observations are essential to confirm whether they are genuinely dark galaxies. With its deep photometric depth, the K-DRIFT observation is promising to provide strong constraints on their existence. Combining the K-DRIFT survey with \ion{H}{i} surveys from the Five-hundred-meter Aperture Spherical radio Telescope and Square Kilometer Array, as well as gravitational lensing maps from the Euclid project, can create a powerful synergy for identifying and investigating dark galaxies. These efforts can be extended to further studies, such as isolated dark galaxies in void regions \cite[see][]{2006A&A...451..817K} and interactions between dark galaxies and luminous galaxies \cite[see][]{2008IAUS..244..235K}. 

\paragraph{Literature-based quantitative references:} The existence and abundance of dark galaxies remain highly uncertain, and \cite{2024ApJ...962..129L} further emphasized that estimating their incidence in simulations is also cautious \cite[see also][]{2026arXiv260104024G}. Observationally, constraints on dark galaxies therefore rely largely on indirect approaches, such as cross-matching \ion{H}{i} sources with optical counterparts. \cite{2025ApJS..279...38K} reported that only $\sim$0.5\% of detected \ion{H}{i} sources can be classified as dark galaxy candidates. Given the substantially deeper photometric depth anticipated for K-DRIFT, this fraction likely represents a conservative upper limit, as faint stellar counterparts may be revealed with deeper imaging. 

\subsubsection{Dwarf Barred Galaxies} \label{sec:targets:dwf:bar}
In the local universe, more than 60\% of disk galaxies are known to host bar structures \cite[][]{1991rc3..book.....D,2015ApJS..217...32B,2019ApJ...872...97L}. These features are often associated with massive, red, gas-poor, and bulge-dominated galaxies, particularly early-type spirals that exhibit strong bars \cite[][]{2008ApJ...675.1141S,2009A&A...495..491A,2009ApJ...692L..34L,2013ApJ...779..162C,2015A&A...580A.116G}. In contrast, late-type or less massive disk galaxies more frequently show weak bars \cite[][]{2019ApJ...872...97L}, indicating that bar strength may be linked to galaxy morphology and evolutionary history. 

Recent observations have begun to reveal a greater diversity of bar morphologies in the low-mass regime. For example, the Galaxy Zoo 2 project \cite[][]{2013MNRAS.435.2835W} reported off-centered bars in low-mass galaxies---reminiscent of Magellanic-type morphologies \cite[][]{2017MNRAS.469.3363K}, highlighting the Large Magellanic Cloud as a representative example \cite[][]{2001AJ....122.1807V,2016AcA....66..149J}. Similarly, \cite{2022MNRAS.516L..24C} reported a lopsided bar in a dwarf galaxy falling into the Virgo Cluster, suggesting a potential role of the environment in shaping bar structures. 

These findings indicate that bar structures in dwarf galaxies are not simply lower-mass analogs of those found in massive systems. Instead, they are likely to arise from a broader range of mechanisms, including spontaneous formation through disk instabilities \cite[][]{1964ApJ...139.1217T,1973ApJ...186..467O} or external triggers such as tidal interactions with companion galaxies or within galaxy clusters \cite[][]{1998ApJ...499..149M,2014MNRAS.445.1339L}. Such external mechanisms may be particularly important in producing the asymmetric and irregular bars observed in less massive or dwarf galaxies. 

Despite their astrophysical significance, dwarf barred galaxies remain underrepresented in statistical studies due to observational limitations. Weak bars are inherently difficult to detect, and dwarf galaxies often exhibit irregular or LSB morphologies that complicate conventional classification. 

The upcoming K-DRIFT survey, with its deep and wide-field imaging capabilities, is well-suited to address this gap. Its high sensitivity will facilitate the detection of faint or weak bars, while its coverage of the southern sky will ensure a large and representative sample. The survey is expected to significantly enhance our understanding of the prevalence, morphology, and origin of bars in the low-mass galaxy population. 

\subsection{Galaxies within Voids} \label{sec:targets:void}
\begin{figure}[t!]
\includegraphics[width=\linewidth]{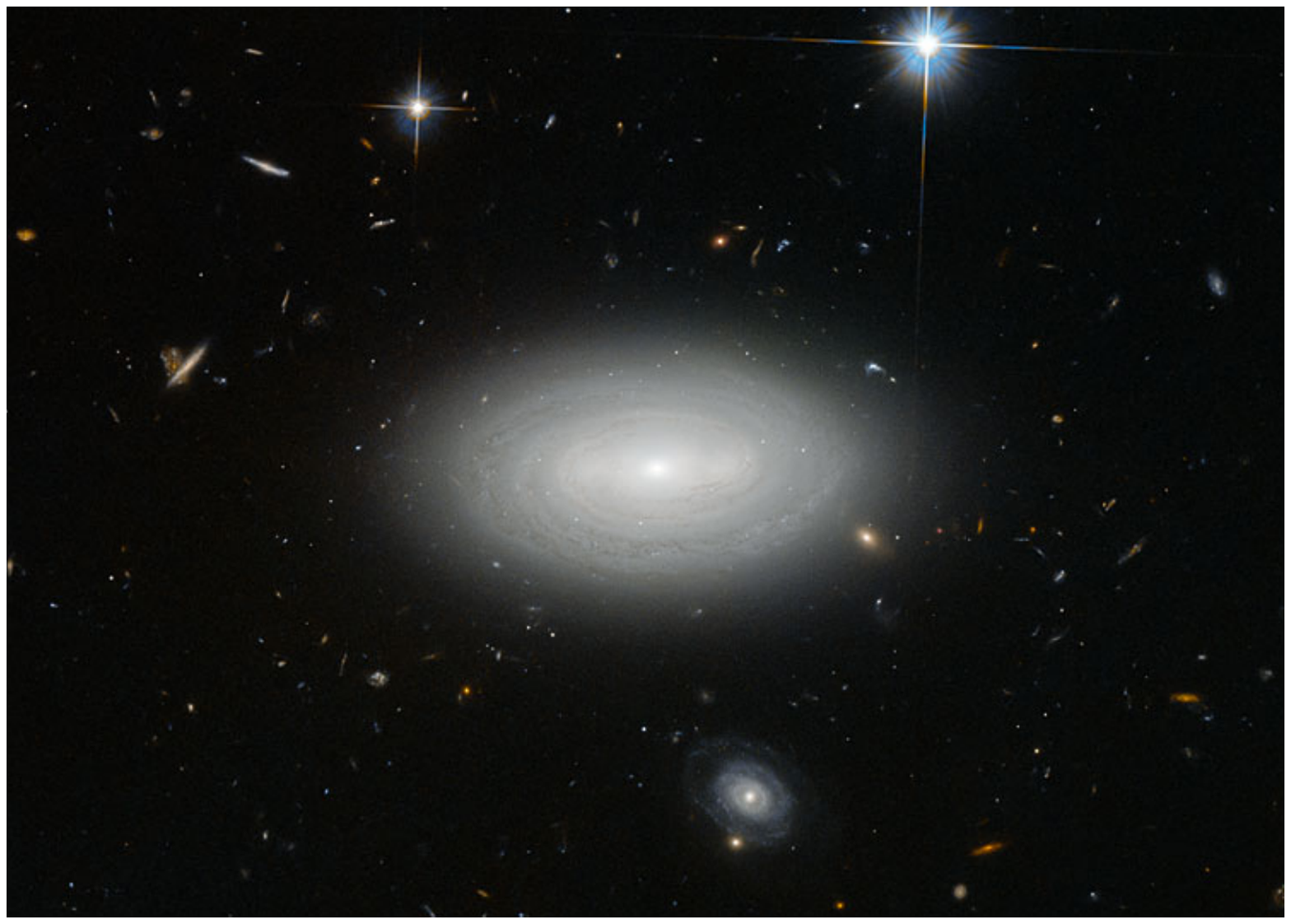}
\centering
\caption{Void galaxy MCG+01-02-015, often referred to as the ``loneliest galaxy'' due to its extreme isolation from large-scale filaments. The FoV of the image is $\sim$$1.8^\prime\times0.8^\prime$. Credit: ESA/Hubble \& NASA and N. Grogin (STScI), Acknowledgement: Judy Schmidt
\label{fig:f8}}
\end{figure}

The large-scale structures of the universe consist of filaments and clusters, which contain numerous galaxies and gas, while the vast spaces surrounding them are called cosmic voids. These voids are the most underdense regions of the universe, but they are not completely empty, as they are often defined as sets of spheres lacking bright galaxies \cite[see][]{2012MNRAS.421..926P,2019MNRAS.482.4329P}. Indeed, even the most mature voids in the local universe contain substructures, including galaxies and diffuse filaments \cite[][]{1996AJ....111.2150S,2004ApJ...607..751H,2012AJ....144...16K,2014MNRAS.440L.106A,2024MNRAS.528..542K}. An excellent example is MCG+01-02-015 (Figure \ref{fig:f8}), a system residing in one of the lowest-density environments. Within this framework, we expect to discover many faint galaxies within voids through the K-DRIFT survey.  

Meanwhile, we might naively assume that void galaxies only undergo secular evolution because of their isolated environments or follow different evolutionary paths compared to galaxies in dense environments. Indeed, void galaxies exhibit several differences from galaxies in dense environments: they are dominated by less massive and bluer galaxies \cite[][]{2004ApJ...617...50R,2005ApJ...620..618H,2012MNRAS.426.3041H,2020MNRAS.491.2496B,2024ApJ...962...58C}, and their specific star formation rate (SFR) tend to be high \cite[][]{2005ApJ...624..571R,2016ApJ...831..118M,2021ApJ...906...97F}.\footnote{On the contrary, some studies suggest minimal differences in SF properties \cite[e.g.,][]{2015ApJ...815...40D,2016MNRAS.458..394B,2022A&A...658A.124D}.} 

%% The "t!" tells LaTeX to put the figure "here" first, at the "top" next
%% and to override the normal way of calculating a float position

Such differences may depend on whether galaxies are central or satellites \cite[see][]{2024MNRAS.528.2822R}. Alternatively, variations in SFR might be attributed to different star formation histories (SFHs). Indeed, \cite{2023Natur.619..269D} showed that the stellar components of void galaxies form more slowly than galaxies in dense environments. One of the possible causes for diverse SFHs is the presence or absence of gravitational interactions that can trigger SF activity. Contrary to the assumption that gravitational interactions are rare for void galaxies, some observational studies have detected signs of interactions \cite[e.g.,][]{2008ApJ...673..715C,2013AJ....145..120B,2016MNRAS.459..754F,2017MNRAS.465.2342C} and mergers \cite[e.g.,][]{2021MNRAS.504.6179E}. 

Recently, \cite{2024MNRAS.528.2822R} showed that void galaxies experience as many mergers as non-void galaxies, but these mergers tend to occur later. If so, numerous tidal features will likely exist in the outskirts of void galaxies. Based on the catalog of \cite{2019MNRAS.482.4329P}, we plan to observe several large voids within 20 Mpc in the southern sky. By utilizing deep imaging survey data, we will attempt to identify tidal features and infer the evolutionary histories of void galaxies. 

\paragraph{Literature-based quantitative references:} \cite{2005ApJ...620..618H} measured the luminosity function of void galaxies using SDSS data and found a faint-end slope of $\alpha \simeq -1.2$, indicating that voids host a substantial population of low-luminosity systems. Focusing on nearby voids, \cite{2019MNRAS.482.4329P} showed that approximately 800 systems with $M_B>-17$ exist. With its significantly deeper imaging, the K-DRIFT survey is expected to extend this census to fainter luminosities. 

\subsection{Galaxies with Interesting Features in the Outskirts} \label{sec:targets:outskirts}
In addition to having distinct tidal structures, the outskirts of galaxies offer benefits for understanding outer structure formation related to SF properties in LSB environments \cite[see][and references therein]{2017ASSL..434.....K}. Generally, the outskirts of galaxies are known to contain few stars and insufficient gas for SF activity. However, they may host extensive outer stellar structures and are not entirely dormant, as they occasionally show SF signals. Due to its vulnerability to stellar feedback, the gas in the outskirts of galaxies may be expelled, leading to a unique SFH. Therefore, investigating the outskirts of galaxies can provide insights into the SF process in LSB environments or the early stages of galaxy formation. This section highlights several types of galaxies with unique and intriguing features, including outer disk structures. 

%% The "t!" tells LaTeX to put the figure "here" first, at the "top" next
%% and to override the normal way of calculating a float position
\begin{figure}[t!]
\includegraphics[width=\linewidth]{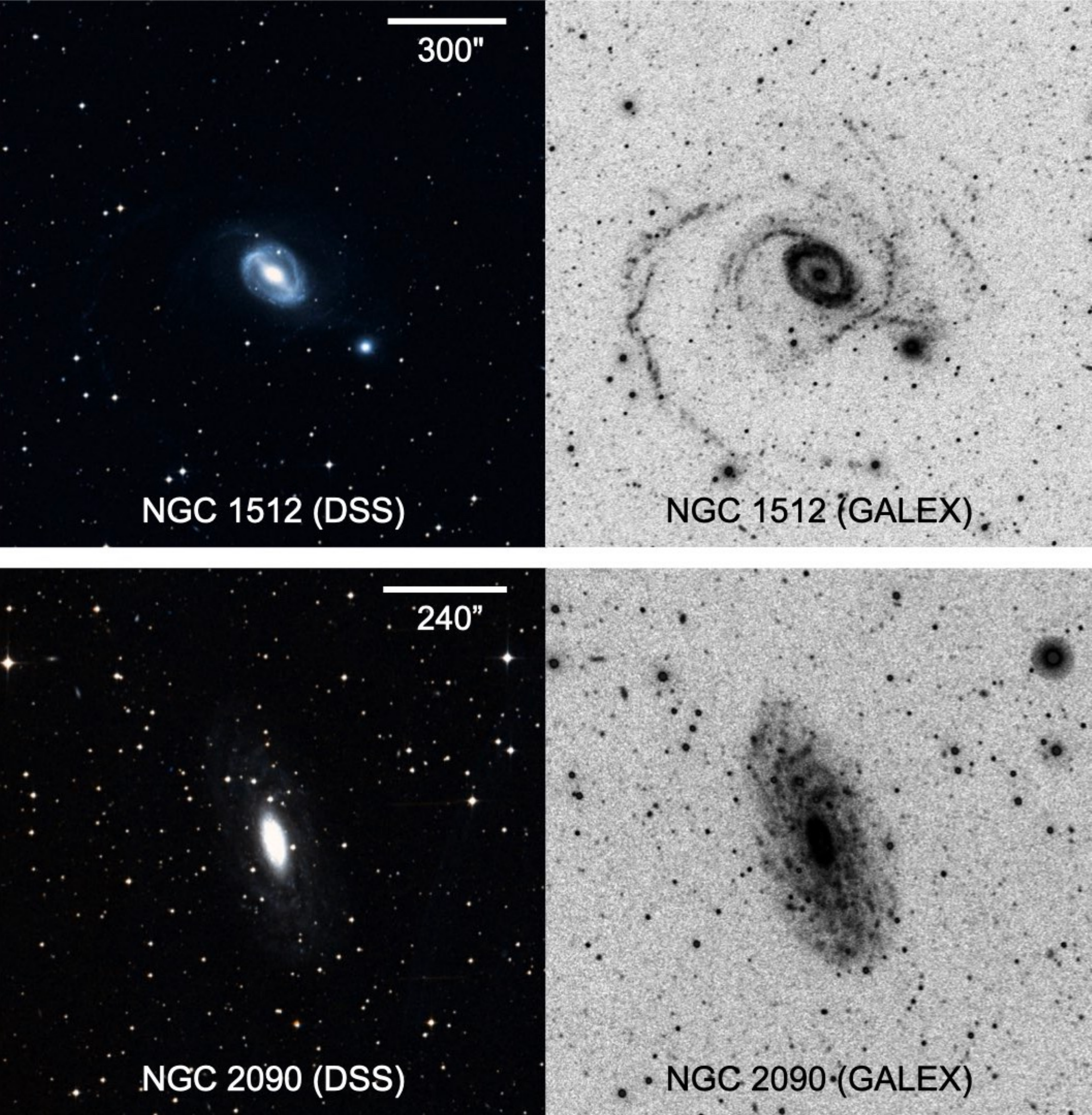}
\centering
\caption{XUV-disk galaxies with Type 1 disk (top) and Type 2 disk (bottom). The UV image (right) prominently shows an enormous star-forming outer disk that is barely visible in the optical image (left).
\label{fig:f7}}
\end{figure}

\subsubsection{Extended Ultraviolet Disk Galaxies} \label{sec:targets:outskirts:xuv}
One of the most intriguing findings of the Galaxy Evolution Explorer (GALEX) satellite \cite[][]{2005ApJ...619L...1M} is the discovery of galaxies with very extended ultraviolet (XUV) disks in their outskirts \cite[][]{2005ApJ...627L..29G,2005ApJ...619L..79T}. These galaxies, dubbed XUV-disk galaxies, are not rare as they account for $\sim$20--40\% of local galaxies \cite[][]{2007ApJS..173..538T,2010MNRAS.405.2791G,2011ApJ...733...74L,2012ApJ...745...34M}. According to \cite{2007ApJS..173..538T}, they can be categorized into two types: Type 1 disks ($\gtrsim$20\% incidence) have distinct UV-bright structures beyond the typical SF threshold, while Type 2 disks ($\sim$10\% incidence) exhibit diffuse UV emission in the outer regions of galaxies (Figure \ref{fig:f7}). 

Earlier, \cite{1989ApJ...344..685K} found that star-forming galaxies exhibit truncated H$\alpha$ profiles in their outer regions, suggesting the existence of a critical gas surface density threshold required for SF activity \cite[see also][]{2001ApJ...555..301M}. In this framework, SF activity in galactic outskirts is considered rare. However, the discovery of XUV-disk galaxies has challenged this view. 

This apparent disagreement may arise from differences in the SF tracers used, as H$\alpha$ and UV emissions reflect SF activity over different timescales. If so, this implies that the observed epoch (or SFH) acts as an important factor in determining the presence of H$\alpha$ disks, suggesting that H$\alpha$ may still be detectable in the outer regions of galaxies. Indeed, \cite{2010MNRAS.405.2791G} reported that some XUV-disk galaxies show extended H$\alpha$ distributions similar to their UV profiles. 

%% The "t!" tells LaTeX to put the figure "here" first, at the "top" next
%% and to override the normal way of calculating a float position
\begin{figure}[t!]
\includegraphics[width=\linewidth]{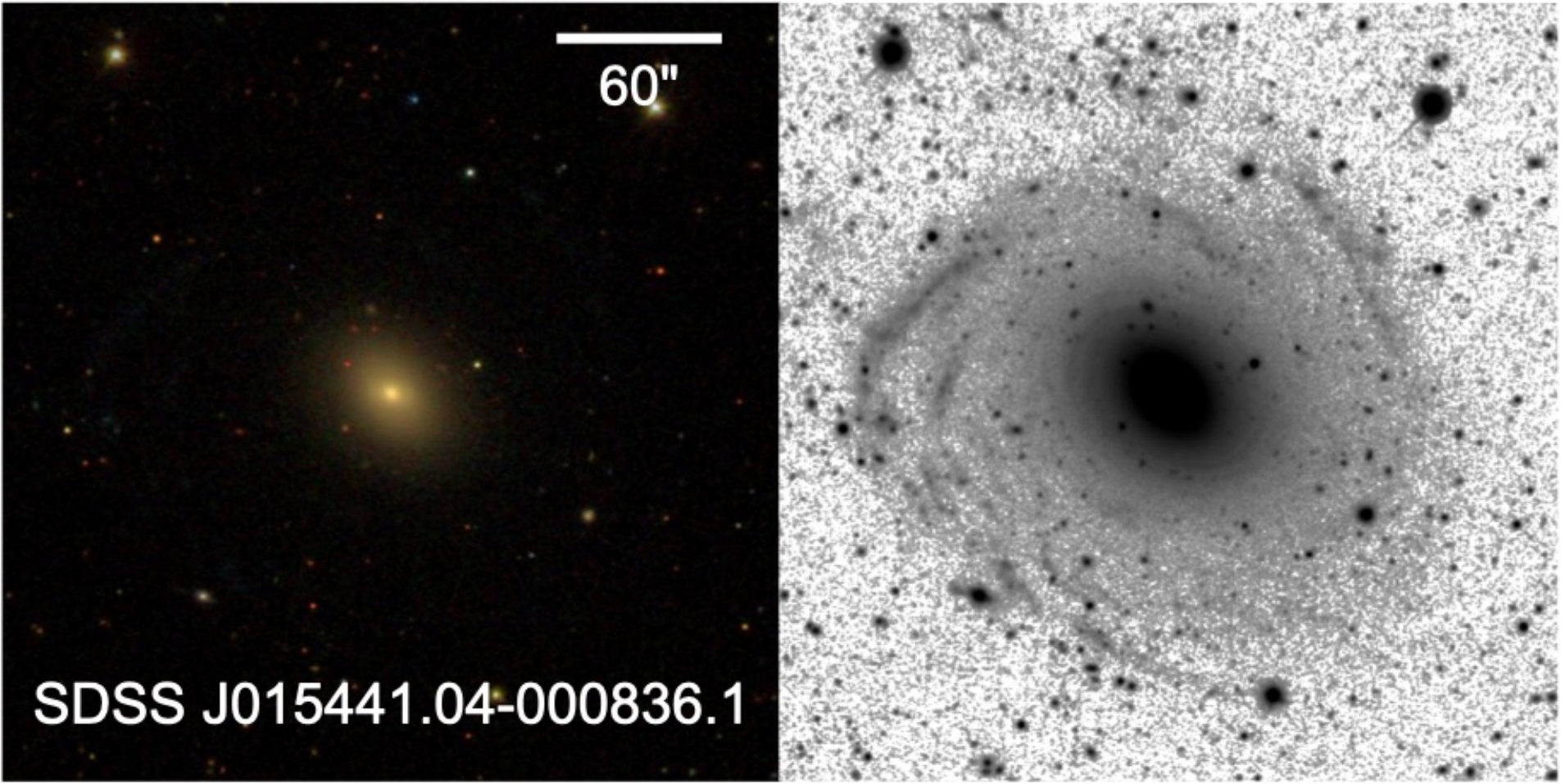}
\centering
\caption{Giant LSB galaxy UGC 1382. A vast and diffuse LSB disk with spiral structures is not discernible in the SDSS single-epoch image (left) but becomes conspicuous in the DESI Legacy Imaging Surveys image (right), which is $\sim$2 mag deeper in surface brightness units.
\label{fig:f9}}
\end{figure}

Consequently, XUV-disk galaxies serve as excellent laboratories for studying SF properties in low gas surface density environments, which are also properties of the LSB regime. This raises an important question: How are stars in the XUV disks formed despite the low gas density? Possible explanations include gas accretion or galaxy interaction, both of which can alter the local gas density and trigger instantaneous SF activity \cite[][]{2007ApJS..173..538T,2011ApJ...733...74L,2012ApJ...745...34M,2021ApJ...918...82B}. 

The deep, wide-field data from the K-DRIFT survey are well-suited for exploring the outskirts of XUV-disk galaxies. Based on existing catalogs, we will conduct revisit observations. Our primary focus is to analyze faint outer disk structures barely visible in optical wavelengths. Additionally, we aim to search for companion galaxies or tidal features that could play a role in the formation of XUV disks. Future observations using an H$\alpha$-narrowband filter will allow us to study SF properties like SFR. 

\subsubsection{Giant LSB Galaxies} \label{sec:targets:outskirts:glsb}
Giant LSB galaxies, characterized by their vast and diffuse LSB disks, have been studied since early on regarding their existence and properties \cite[e.g.,][]{1987AJ.....94...23B,1997AJ....114.1858P,2007AJ....133.1085B,2016ApJ...826..210H,2019MNRAS.489.4669S,2021MNRAS.503..830S,2023MNRAS.520L..85S}. Their LSB disks are typically fainter than $\mu_r\sim26$ mag arcsec$^{-2}$ and extend up to $\sim$100 kpc in radius. Because giant LSB galaxies are observed with overwhelmingly bright central regions compared to disks, some have likely been misclassified as passive ETGs: excellent examples include Malin 1 \cite[][]{2015ApJ...815L..29G} and UGC 1382 \cite[][]{2016ApJ...826..210H}. As shown in Figure \ref{fig:f9}, a single-epoch SDSS image shows only the central region due to its shallow photometric depth, making the galaxy appear ETG-like. However, the deep co-added image from the DESI Legacy Imaging Surveys reveals the LSB disk with spiral structures. 

Giant LSB galaxies are considered very rare objects, with a volume density of $\sim$$4\times10^{-5}$ Mpc$^{-3}$ \cite[][]{2023MNRAS.520L..85S} as it is difficult for such massive ($M_\star\gtrsim10^{11}$ $M_\odot$) galaxies to retain a (LSB) disk. Various scenarios have been proposed to explain the origin of these structures, including major collision-induced perturbations \cite[e.g.,][]{2008MNRAS.383.1223M,2018MNRAS.481.3534S}, accretion of tidally disrupted dwarf galaxies \cite[e.g.,][]{2006ApJ...650L..33P,2016ApJ...826..210H}, interactions with companions \cite[e.g.,][]{2010MNRAS.406L..90R}, or multiple formation pathways \cite[see][]{2021MNRAS.503..830S}. On the contrary, some studies have shown that giant LSB galaxies can be formed without external influences, emerging naturally in isolated environments \cite[e.g.,][]{1992ApJ...388L..13H,2001MNRAS.328..353N,2014MNRAS.437.3072K,2016A&A...593A.126B,2019MNRAS.489.4669S}. 

We anticipate that the K-DRIFT survey will discover at least 100 giant LSB galaxies in the local universe. Its deep photometric depth will allow for the detection of galaxies with extremely faint ($\mu_r\gtrsim28$ mag arcsec$^{-2}$) disks, if such galaxies exist. Using this sample, we will analyze the SB profile and color gradient, and additionally, we will closely examine whether there are tidal features or companion galaxies. These analyses will help clarify the origin and formation mechanisms of giant LSB galaxies. In this regard, investigating their environments and/or \ion{H}{i} distributions jointly will be an effective approach. 

%% The "t!" tells LaTeX to put the figure "here" first, at the "top" next
%% and to override the normal way of calculating a float position
\begin{figure}[t!]
\includegraphics[width=\linewidth]{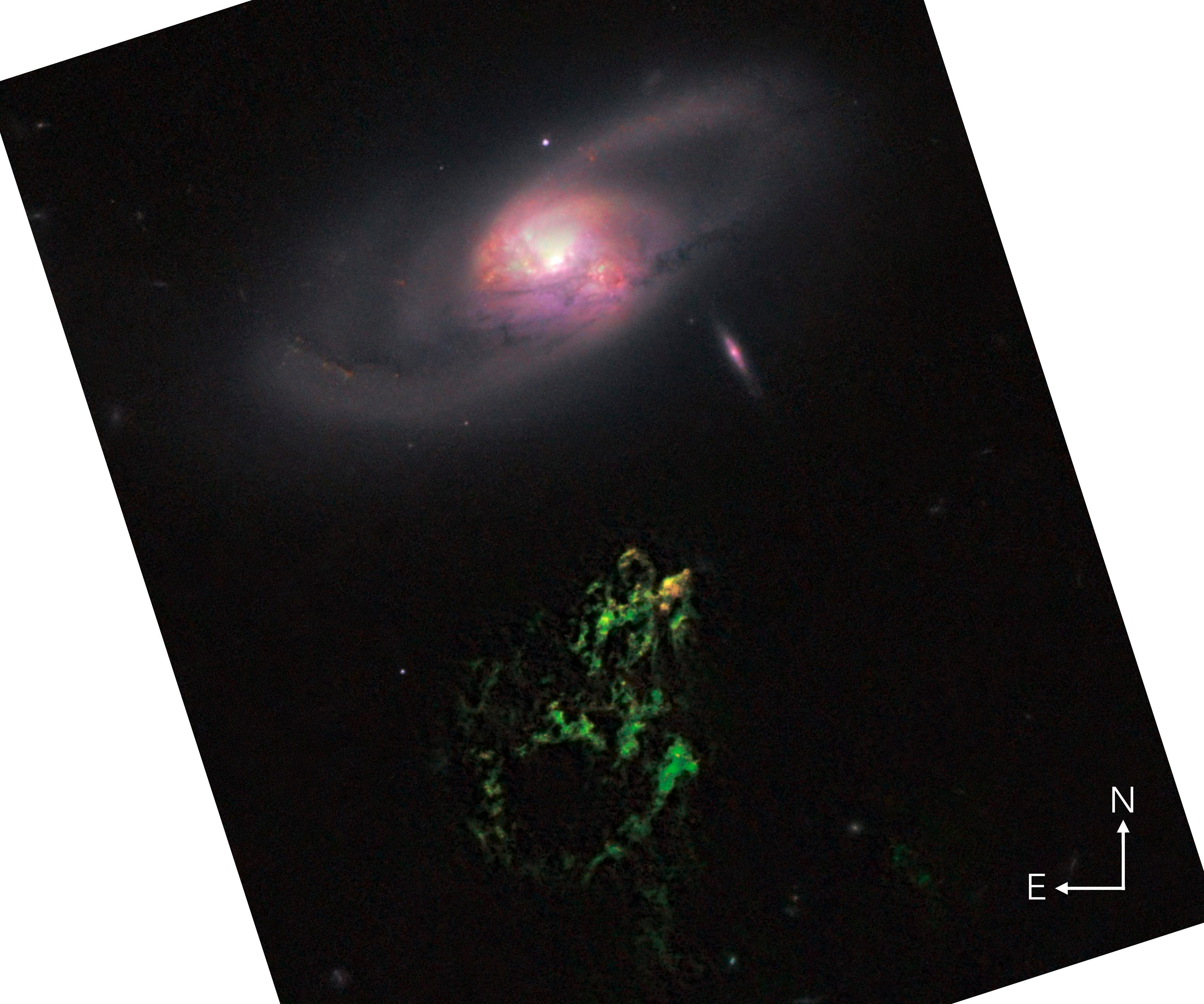}
\centering
\caption{Hanny's Voorwerp (greenish), located south of IC 2497. The image has a FoV of $\sim$$0.8^\prime\times1.0^\prime$ and is rotated by $\sim$18$^\circ$. Credit: NASA, ESA, William Keel (University of Alabama, Tuscaloosa), and the Galaxy Zoo team
\label{fig:f10}}
\end{figure}

Meanwhile, we tentatively propose that giant LSB galaxies are evolutionarily connected to XUV-disk galaxies. Indeed, giant LSB galaxies often exhibit signs of SF activity in their LSB disks \cite[][]{2008ApJ...681..244B}, and the two galaxies mentioned earlier are classified as XUV-disk galaxies \cite[see][]{2008ApJ...681..244B,2016A&A...593A.126B,2016ApJ...826..210H}. This suggests that (at least) Type 1 XUV-disk galaxies might be a subcategory of giant LSB galaxies. Further comparative studies focusing on their SFHs and gas properties will help to better understand their nature. 

\subsubsection{Analogs of Hanny's Voorwerp} \label{sec:targets:outskirts:hanny}
``Hanny's Voorwerp'' is a fascinating nebula feature discovered in 2007 by Hanny van Arkel, one of the volunteers in the Galaxy Zoo project.\footnote{\url{https://www.zooniverse.org/projects/zookeeper/galaxy-zoo}} As shown in Figure \ref{fig:f10}, this gas cloud is approximately 25 kpc away from the adjacent galaxy IC 2497 and emits a distinctive green light due to strong [\ion{O}{iii}] emission lines \cite[][]{2009MNRAS.399..129L}. Additionally, the regions facing IC 2497 exhibit signs of ongoing SF activity.

The gas cloud is believed to be ionized by strong UV and X-ray emissions from the active galactic nucleus (AGN) of IC 2497. However, the AGN does not appear bright, suggesting that it may either be heavily obscured or have recently experienced a sudden shutdown \cite[][]{2009MNRAS.399..129L}. The latter scenario suggests that energy released in the recent past is still propagating outward and ionizing the distant gas cloud. In addition, \cite{2012AJ....144...66K} revealed fine structures emitting H$\alpha$ radiation through HST observations and suggested that AGN outflows may have interacted with these regions, triggering SF activity. For this reason, Hanny's Voorwerp is often referred to as a quasar ionization echo and is also considered evidence of a rapid change in AGN activity. 

%% The "t!" tells LaTeX to put the figure "here" first, at the "top" next
%% and to override the normal way of calculating a float position
\begin{figure*}[t!]
\includegraphics[width=\linewidth]{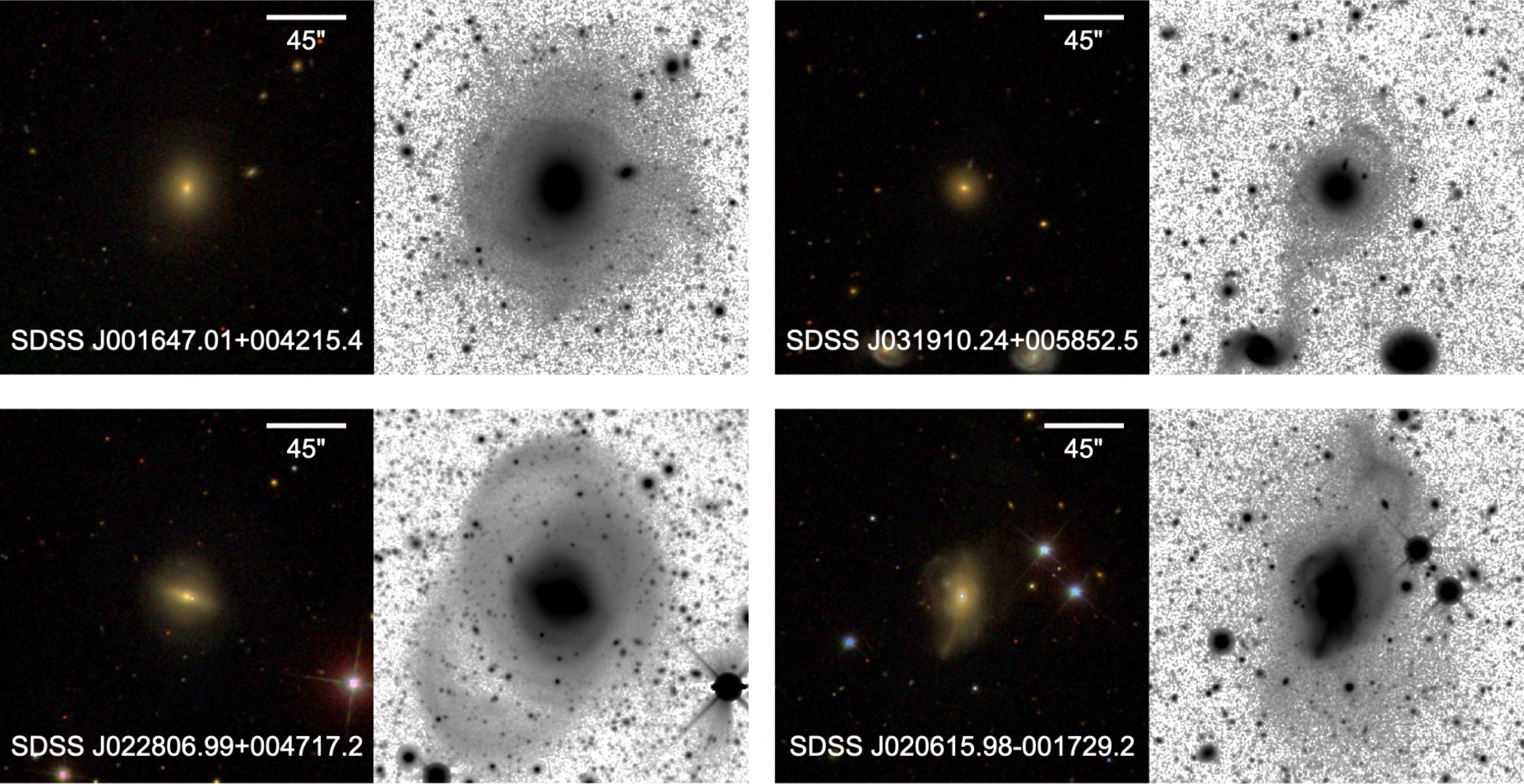}
\centering
\caption{Comparison between the SDSS single-epoch images (color) and the DESI Legacy Imaging Surveys images (grayscale), which are $\sim$2 mag deeper in surface brightness units. Tidal features not discernible in the single-epoch images are clearly identified in the deep images, including (from top left to bottom right) shell, tidal tail, stellar stream, and AGN-involved structures. 
\label{fig:f11}}
\end{figure*}

Meanwhile, where do the distant gas clouds come from? Through \ion{H}{i} kinematic analysis, \cite{2009A&A...500L..33J} proposed that these gas clouds likely originated from tidal interaction. This hypothesis is further supported by \cite{2012MNRAS.420..878K}, who reported that 14 out of 19 galaxies with AGN-ionized clouds are in the phase of interacting or merging, suggesting that the gas clouds mainly originated in tidal tails. Additionally, ongoing studies continue to strengthen this interpretation, providing additional evidence for a connection to dynamical interaction \cite[e.g.,][]{2013ApJ...773..148S,2015AJ....149..155K,2024MNRAS.530.1624K}. 

From this perspective, the deep optical imaging data from the K-DRIFT survey can play a crucial role by enabling the detection of even the faintest tidal features. If ring or bridge structures extending from host galaxies to glowing gas clouds are identified, they would serve as strong evidence for dynamical interactions. Meanwhile, the discovery of Hanny's Voorwerp analogs is inherently biased toward the northern hemisphere sky because it relied heavily on the Galaxy Zoo project, which is primarily based on SDSS data. We expect that our future observations using H$\alpha$ and/or [\ion{O}{iii}] filters will allow us to discover Hanny's Voorwerp analogs in the southern sky. Furthermore, as described in Section \ref{sec:impact:agn}, we also plan to investigate the connection between tidal features and AGN activity. 

\subsection{Enhancing the Detection and Characterization of LSB Objects using ML Techniques} \label{sec:targets:ml}
Because of their faint and diffuse nature, LSB galaxies/features have long posed challenges in their detection and characterization. However, recent advancements in ML techniques have opened promising pathways for improving the identification and understanding of LSB features in the universe. In this section, we discuss the implications of utilizing ML methods to analyze the K-DRIFT survey data in the context of LSB object research. 

Many studies have demonstrated that ML algorithms are invaluable tools for addressing the complexity of LSB galaxies. For example, deep learning and convolutional neural networks have proven effective in distinguishing LSB galaxies from artifacts \cite[e.g.,][]{2021A&C....3500469T,2025ApJ...995..178P,2025A&A...701L...9P} and in classifying dwarf galaxies or tidal features from multi-band imaging data \cite[e.g.,][]{2019MNRAS.483.2968W,2021OJAp....4E...3M,2024MNRAS.534.1459G}. Additionally, the application of self-supervised representation learning has enabled the detection of tidal features while saving time and effort \cite[e.g.,][]{2023mla..confE..11D}. Furthermore, \cite{2023MNRAS.519.4735S} introduced an innovative approach using ML and neural networks to study the Galactic cirrus filaments, which are obstacles in LSB study. This provided an opportunity to better understand the Galactic cirrus and allowed us to correctly investigate LSB objects. Recently, \cite{2024A&A...687A..24M} assessed several ML methods to classify merging systems using various datasets, including simulation and observation data. They showed that the ML methods perform reasonably well but are partly limited by redshift and imaging quality. 

When delving into the extensive K-DRIFT survey data to explore the LSB universe, ML algorithms provide a powerful means of efficiently identifying and characterizing LSB objects. Employing a combination of simulation data and validated observational data will serve as practical training for these algorithms. This approach will enable the identification of even the most subtle LSB signatures, which might elude traditional analysis methods such as visual inspection. 

\section{Role of LSB Observations in Understanding Galaxy Merger and Its Evolutionary Impact} \label{sec:impact}
In the $\Lambda$CDM universe, galaxy mergers play a critical role in the formation and evolution of galaxies \cite[][]{1996MNRAS.283.1361B,2007MNRAS.375....2D,2013MNRAS.433.2986W}. For example, multiple mergers serve as a primary pathway for the growth of massive galaxies \cite[][]{2017ApJ...834...73Y,2022ApJ...936...22Y}. Gas-rich mergers can also trigger intense SF activity, depleting gas and producing quiescent remnants within $\sim$1 Gyr \cite[][]{1996ApJ...464..641M,2005ApJ...620L..79S,2015MNRAS.451.2933B}. At the same time, AGN feedback triggered during these interactions may act to suppress further SF events \cite[][]{2005ApJ...625L..71H,2005ApJ...620L..79S}. Beyond these effects, mergers drive significant structural and dynamical transformations, including the concentration of stellar light in ETGs \cite[][]{2006MNRAS.369..625N,2009ApJ...691.1828P,2013MNRAS.429.2924H}, the expansion of galaxy sizes through dry minor mergers \cite[][]{2013MNRAS.428..641O,2017ApJ...834...73Y}, and the reshaping of stellar kinematics in merger remnants \cite[][]{2017MNRAS.464.3850L,2017MNRAS.468.3883P,2022ApJ...925..168Y}. 

Numerical simulations have provided more detailed insights into the impact of galaxy mergers on the properties of galaxies. For instance, several studies have demonstrated that mergers influence the internal stellar kinematics of galaxies \cite[e.g.,][]{2009MNRAS.397.1202J,2011MNRAS.416.1654B,2017ApJ...837...68C,2017ApJ...850..144E}. Additionally, analyses of stellar populations have revealed that dry mergers tend to flatten radial metallicity profiles, while gas-rich mergers can produce steep negative metallicity and positive age gradients \cite[e.g.,][]{2004MNRAS.347..740K,2009A&A...499..427D,2016ApJ...833..158C,2017MNRAS.471.3856T}. 

Unlike simulations that allow for temporal tracking, observational studies provide only snapshots of the universe. This limitation makes it challenging to directly determine the impact of past mergers on current galaxy properties or to validate simulation results. However, tidal features offer a valuable solution by serving as historical tracers of merger events, extending the temporal scope of observations to the past few Gyr, as described in Section \ref{sec:targets:tidal}. Achieving deep photometric depths in imaging data is critical for reliably detecting these features. Figure \ref{fig:f11} highlights this importance, showing how deep imaging can reveal merger histories even in galaxies that appear relaxed at first glance. 

Meanwhile, diversity in datasets and their corresponding photometric depths can lead to inconsistent findings. For instance, \cite{2010ApJS..186..427N} reported that only $\sim$3--6\% of ETGs exhibit tidal or disturbed features in the single-epoch SDSS images. This fraction increased up to $\sim$10--25\% in the Stripe 82 deep coadded images \cite[][]{2018ApJ...857..144H,2020ApJ...905..154Y,2022ApJ...925..168Y,2023ApJ...946...41Y}. Additionally, \cite{2005AJ....130.2647V} identified tidally disturbed features in $\sim$70\% of nearby ETGs using even deeper images. The discrepant results are attributed to differences in the detectable SB limits of each dataset, which are approximately 25, 27, and 29 mag arcsec$^{-2}$ in turn, respectively. 

Hence, the deep and wide-field images of the K-DRIFT survey will facilitate systematic investigations of tidal features, greatly aiding in studying the impact of galaxy mergers on galaxy properties. This section presents several case studies of galaxy mergers and their evolutionary implications, with a focus on the presence of tidal features. The spatially resolved properties of the host galaxies discussed here are derived from integral field unit (IFU) spectroscopic data, which are expected to synergize with future work. 

\subsection{Tendency in Photometric Parameters of ETGs} \label{sec:impact:phot}
Many previous studies have examined the correlation between the presence of tidal features and photometric properties of ETGs, especially their colors. For example, \cite{1992AJ....104.1039S} found that ETGs with bluer colors at a given luminosity exhibit more merging structures. In addition, \cite{2010ApJ...714L.108S} showed that tidal features are more frequently detected in young ETGs with bluer colors or past SF signals in emission lines than in old and quiescent ETGs. Similarly, \cite{2009AJ....138.1417T} and \cite{2011MNRAS.411.2148K} found that morphologically disturbed ETGs tend to have bluer optical colors than ETGs without disturbances. These results suggest that recent mergers are closely related to blue (young) stellar populations in ETGs. 

\begin{figure}[t!]
\includegraphics[width=\linewidth]{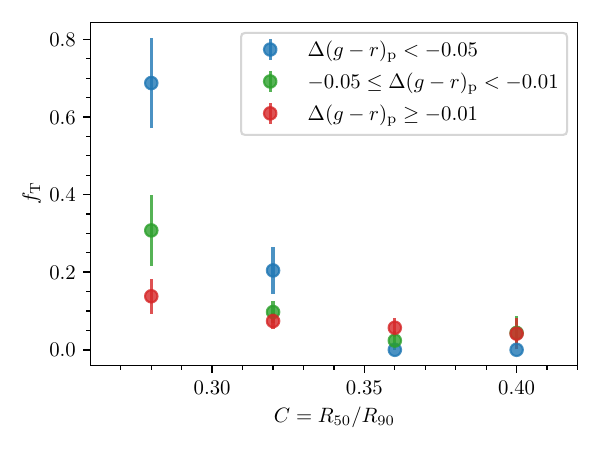}
\centering
\caption{Fraction of ETGs with tidal features ($f_\mathrm{T}$) as a function of the concentration index ($C$). The samples are divided into three color bins of $\Delta(g-r)_\mathrm{P}$. More concentrated and bluer ETGs have higher $f_\mathrm{T}$ than those with redder color and less concentrated light profiles. This figure is a reproduced version of Figure 9 in \cite{2020ApJ...905..154Y}.
\label{fig:f12}}
\end{figure}

Recently, \cite{2020ApJ...905..154Y} examined the correlation between the fraction of ETGs with tidal features ($f_\mathrm{T}$) and their photometric properties using 650 samples at $z\leq0.055$ in the SDSS Stripe 82 data. Figure \ref{fig:f12} shows one of the key results: tidal features are more frequent in bluer and concentrated ETGs than in redder and less compact counterparts. In particular, compact ETGs with blue cores have significantly higher $f_\mathrm{T}$ than those with red cores. They also found that ETGs with dust lanes have much higher $f_\mathrm{T}$ than those without dust lanes. These results also suggest that blue and concentrated ETGs, as well as their dust lanes, are involved in recent mergers. 

%% The "t!" tells LaTeX to put the figure "here" first, at the "top" next
%% and to override the normal way of calculating a float position

The deep images of the K-DRIFT survey allow us to ensure much larger galaxy samples. Additionally, we can detect extremely faint tidal features, likely debris from mergers with small galaxies or that occurred long ago. With much more systematic investigations, we will verify whether the results are consistent with previous works. Furthermore, we will perform more advanced analyses, including classifying the tidal features and measuring their photometric properties. These studies will further expand our understanding of how galaxy mergers affect the evolution of galaxies. 

\subsection{Tendency in Spectroscopic Parameters of ETGs} \label{sec:impact:spec}
Extensive IFU spectroscopic surveys \cite[e.g.,][]{2011MNRAS.413..813C,2012A&A...538A...8S,2015ApJ...798....7B,2016SPIE.9908E..1FB,2021MNRAS.505..991C} enable the study of spatially resolved kinematics and stellar population for a large number of galaxies with diverse morphologies. For example, \cite{2021ApJ...922..249Y} extracted stellar rotation curves of nearly 2000 galaxies from the SDSS Mapping Nearby Galaxies at Apache Point Observatory (MaNGA) survey \cite[][]{2015ApJ...798....7B} and found that only massive LTGs and S0 galaxies have a flat rotation curve at large radii, which is reproduced in numerical simulations \cite[][]{2025ApJ...982...11J}. In addition, several studies of radial stellar population profiles of ETGs commonly found that ETGs tend to exhibit slightly positive or flat age profiles and negative metallicity profiles \cite[e.g.,][]{2017MNRAS.466.4731G,2020A&A...644A.117L,2021MNRAS.508.4844N}. 

\begin{figure}[t!]
\includegraphics[width=\linewidth]{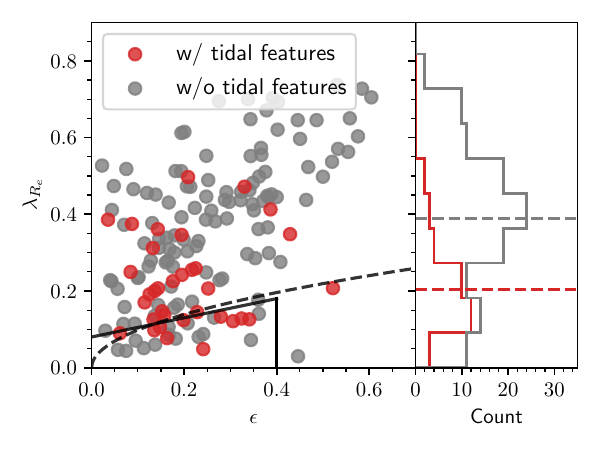}
\centering
\caption{Left: the distribution on the specific angular momentum ($\lambda_{R_e}$) versus ellipticity ($\epsilon$) plane, with ETG samples divided into two groups: those with and without tidal features. The black dashed and solid lines represent the slow rotator boundaries defined by \cite{2011MNRAS.414..888E} and \cite{2016ARA&A..54..597C}, respectively. Right: the one-dimensional $\lambda_{R_e}$ distributions for the two groups, with horizontal dashed lines indicating the median value for each group. This figure is a reproduced version of Figure 4 in \cite{2023ApJ...946...41Y}.
\label{fig:f13}}
\end{figure}

Combining spatially resolved galaxy properties obtained from IFU spectroscopic data with tidal features in deep images makes it feasible to investigate more specifically how galaxy mergers affect the kinematics and stellar population distributions of galaxies. For example, \cite{2022ApJ...925..168Y} examined the distribution of specific angular momentum of ETGs with and without tidal features using MaNGA data and SDSS Stripe 82 coadded images. They found that the specific angular momentum is lower in ETGs with tidal features than those without tidal features (Figure \ref{fig:f13}). Moreover, \cite{2023ApJ...946...41Y} found that the stellar population profiles of ETGs with tidal features have less negative metallicity gradients and more positive age gradients than those without tidal features. 

%% The "t!" tells LaTeX to put the figure "here" first, at the "top" next
%% and to override the normal way of calculating a float position

As mentioned in the previous section, the deep images of the K-DRIFT survey will provide a large sample of galaxies, including those with extremely faint tidal features. By combining them with extensive IFU survey data, we plan to scrutinize the impact of galaxy mergers on the kinematics and stellar population distributions of galaxies. Meanwhile, the formation epoch and mechanism of the outer parts of ETGs are likely different from those governing the inner parts \cite[see][]{2009ApJ...699L.178N,2010ApJ...725.2312O}. For samples with IFU data covering up to approximately 3--5 half-light radii, we expect to obtain novel insights, as the galaxy outskirts and tidal features will be directly probed. 

\subsection{Origin and Evolution of Star-Forming ETGs} \label{sec:impact:sf-etg}
ETGs are generally considered quiescent in terms of SF activity \cite[][]{2012MNRAS.424..232W,2019ApJ...878...69L}. However, some ETGs have been revealed to exhibit strong H$\alpha$ emissions, similar to those typically seen in LTGs \cite[e.g.,][]{2004ApJ...601L.127F}. These galaxies, often called star-forming ETGs, account for $\sim$4--7\% of ETGs in the local universe \cite[][]{2019ApJ...878...69L,2022MNRAS.509..550J,2023ApJ...953...88L}. 

Star-forming ETGs typically contain gas in the central regions and often exhibit kinematic misalignments between the gas and stellar components \cite[][]{2023ApJ...953...88L}. This misalignment is likely related to external gas supply, including galaxy mergers or gas accretion from the cosmic web \cite[see][]{2006MNRAS.366.1151S,2016MNRAS.463..913J,2022ApJ...934L..35C,2023ApJ...953...88L}. Indeed, faint merger features have been detected in the outskirts of some of these galaxies \cite[e.g.,][]{2022MNRAS.509..550J,2023A&A...678A..10G}. For this reason, star-forming ETGs have been suggested as an outcome of the rejuvenation of passive ETGs \cite[][]{2009MNRAS.398.1651H,2022MNRAS.513..389R}. 

However, an analysis of the properties of star-forming ETGs, such as stellar mass and environment, revealed that they share more similarities with LTGs than with typical ETGs \cite[][]{2023ApJ...953...88L}. This finding suggests that star-forming ETGs may represent an early stage in which less-massive galaxies or those in low-density environments morphologically transform into ETGs \cite[see][]{2022MNRAS.515..213P,2023ApJ...953...88L}. Supporting this claim, many simulation studies have demonstrated that misaligned gas can contribute to bulge formation or even drive morphological transformation \cite[e.g.,][]{2009MNRAS.396..696S,2011MNRAS.418.2493P,2012MNRAS.423.1544S,2015MNRAS.450.2327Z,2019ApJ...883...25P}. 

In this respect, the extensive and deep images obtained from the K-DRIFT survey can provide valuable insight into the origin of star-forming ETGs and the morphological transformation process from LTGs to ETGs. Specifically, we plan to distinguish and examine star-forming ETGs with and without merger features. We expect this statistical approach to reveal various correlations between the significance of merger features and host galaxy properties, such as color, Sersic index, and SFR. 

\subsection{Formation of Elliptical Galaxies in Isolated Environments} \label{sec:impact:isolated}
Galaxies reside in a wide range of environments, from fields to clusters, and their evolutionary aspects can vary significantly depending on these surroundings. For example, massive ETGs in cluster-like environments tend to be larger compared to those in fields \cite[][]{2013MNRAS.435..207L,2014MNRAS.441..203D,2017ApJ...834...73Y}. In addition, disk galaxies in interacting clusters seem to experience bar formation and SFR enhancement \cite[][]{2019NatAs...3..844Y,2020ApJ...893..117Y}. Most notably, denser environments favor being dominated by early-type or quiescent galaxies \cite[][]{1980ApJ...236..351D,2009ApJ...700..791H}. 

Meanwhile, quiescent elliptical galaxies have also been discovered in very isolated environments, with no neighbors within a projected distance of $\sim$1 Mpc \cite[][]{2004AJ....127.1336S,2016A&A...588A..79L,2022PASA...39...31U}. Since elliptical galaxies are typically thought to be formed through mergers, several studies have proposed the possibility that these galaxies may have grown by cannibalizing nearby galaxies, ultimately becoming the only galaxies in the regions \cite[e.g.,][]{1999ApJ...514..133M,2004MNRAS.354..851R}. 

In this regard, numerical simulations have shown that elliptical galaxies in isolated environments are likely to have experienced mergers more recently compared to those in dense environments \cite[e.g.,][]{2010MNRAS.405..477N,2017ApJ...834...73Y}. Indeed, several observational studies have identified tidal features around elliptical galaxies in isolated environments \cite[e.g.,][]{1996MNRAS.282..149R,2001AJ....121..808C,2004MNRAS.354..851R,2008ApJ...674..784P}. Recently, Yoon et al. (in preparation) found circumstantial evidence of mergers even in low-mass quiescent galaxies with a stellar mass of $\lesssim$10$^{10}$ $M_\odot$, in contrast to those in denser environments, which are predominantly formed by environmental effects such as ram pressure, strangulation, and harassment. 

Since isolated environments, such as voids, are one of the main targets of the K-DRIFT survey, we anticipate obtaining a large sample of galaxies in these regions. Our goal is to conduct an in-depth investigation to explore whether the formation of elliptical galaxies in isolated environments is linked to recent mergers, as indicated by the presence of tidal features. While massive galaxies, which are likely to have undergone multiple mergers, will be our primary targets, we will also search for direct evidence of mergers around low-mass quiescent galaxies. 

\subsection{Merger-Induced SFR Enhancement in LTGs} \label{sec:impact:sfr_ltg}
Many simulation works have illustrated that tidal interactions during mergers can compress gas and stimulate SF activity \cite[e.g.,][]{2004MNRAS.350..798B,2005ApJ...620L..79S,2009ApJ...694L.123K,2009PASJ...61..481S,2019MNRAS.485.1320M}, especially during close encounters and coalescence phases. However, the triggered SF may rapidly cease after the coalescence phase due to subsequent AGN and/or SN feedback \cite[][]{2003MNRAS.339..289S,2005ApJ...620L..79S,2008ApJS..175..356H}. 

Similarly, several observational studies have reported increased SFR in galaxies undergoing mergers or those close to each other \cite[e.g.,][]{2009ApJ...691.1828P,2011A&A...535A..60H,2012MNRAS.426..549S,2013MNRAS.435.3627E,2015MNRAS.454.1742K,2018ApJ...868...46S,2019A&A...631A..51P}. However, detailed observational research on the efficiency and duration of SFR enhancement is still lacking. To better understand the role of mergers, it is necessary to examine their significance, which can be achieved by inferring the progenitors involved in mergers. 

Major mergers can cause violent changes not only in SFR but also in morphology, e.g., driving transitions from star-forming to quiescent and/or from late-type to early-type \cite[see][]{1988ApJ...331..699B,1992ApJ...399L.117H,1999ApJ...523L.133N,2006MNRAS.369..625N}. To ensure robust statistical analyses of SFR enhancement, it is essential to control the sample to avoid contamination. Therefore, we will focus only on LTGs with tidal features, indicative of minor mergers a few Gyr ago \cite[][]{2014A&A...566A..97J,2019A&A...632A.122M,2020ApJ...905..154Y}, but without significant morphological transformations. The K-DRIFT survey data will enable us to reliably identify LTGs with tidal features and assess their significance. Subsequently, we will test SFR enhancement using multiple archival IFU datasets. 

\subsection{Merger-Triggered AGN Activity} \label{sec:impact:agn}
AGNs are the central regions of galaxies where supermassive black holes (SMBHs) accrete surrounding material, releasing extremely high-energy radiation. The most luminous AGNs, called quasars, emit energy more than 100 times stronger than normal galaxies like MW \cite[][]{2015Natur.518..512W}. The energetic feedback from AGNs can significantly impact their host galaxies \cite[see][and references therein]{2012ARA&A..50..455F}. In this way, SMBHs and their host galaxies are believed to co-evolve, as supported by the tight scaling relation observed between them \cite[see][]{2013ARA&A..51..511K}. Consequently, studying AGNs is crucial for understanding both galaxy evolution and SMBH growth. 

Galaxy mergers are one of the key mechanisms that can trigger AGN activity by driving gas infall into SMBHs \cite[][]{2005ApJ...620L..79S,2008ApJS..175..356H,2015ApJ...804...34H}. In particular, gas-rich major mergers are often associated with the most luminous AGNs \cite[][]{2006ApJS..163....1H,2012ApJ...758L..39T,2016ApJ...822L..32F}. On the contrary, some observational studies suggest that major mergers account for only a small fraction of AGN-triggering events, with most AGNs being activated by secular processes \cite[e.g.,][]{2011ApJ...727L..31S,2012MNRAS.425L..61S}. In such cases, bar structures of galaxies may act as a pathway of gas inflow toward the central regions, fueling AGN activity \cite[][]{1989A&A...217...66A,2002ApJ...567...97L,2002MNRAS.329..502M,2004ApJ...600..595R}. 

To address the debate about AGN-triggering mechanisms, tidal features can serve as an effective probe. Using the K-DRIFT survey data together with IFU data, we will categorize AGN-host galaxies as either exhibiting or lacking tidal features and analyze their properties. This includes kinematics and stellar populations of host galaxies, as well as bolometric luminosity and Eddington ratio for AGNs. Given that the typical lifetimes of tidal features are comparable to the duration of AGN activity \cite[see][]{2008ApJS..175..356H}, comparing the properties of these two categories will provide an opportunity to examine the differences arising from distinct AGN-triggering mechanisms. Additionally, we will compare AGN-host galaxies with inactive galaxies that exhibit similar features. This analysis will shed light on how AGN feedback impacts host galaxies during mergers while also offering insights into the duration of AGN activity and the lifespan of tidal features. 

\section{Concluding Remarks} \label{sec:con}
The K-DRIFT project aims to advance LSB astronomy through deep, wide-field imaging designed to detect features as faint as 29–30 mag arcsec$^{-2}$. Its off-axis freeform optical design and rolling-dither strategy enable uniform ultra-deep imaging, allowing systematic studies of faint objects in the southern sky. These capabilities fill a gap left by conventional observing systems, which lack sufficient photometric depth.

With K-DRIFT G1, we will uncover various LSB structures, such as stellar streams, shells, tidal tails, and diffuse stellar halos, thus illuminating galaxy mass assembly histories and environmental effects on galaxy evolution. The survey will also expand the census of dwarf satellites, UDGs, compact ellipticals, and possible DMDGs or dark-galaxy candidates in the nearby universe. Additionally, deep imaging of galaxy outskirts will reveal giant LSB disk structures, offering new insights into the formation of outer structures in galaxies. 

The scientific impact of K-DRIFT will be enhanced through synergy with \ion{H}{i} mapping, IFU spectroscopy, and upcoming large-scale surveys, enabling multi-wavelength studies of baryons and dark matter in the faint regime. By providing the most comprehensive census of LSB features in the local universe, K-DRIFT will establish a vital observational foundation for testing small-scale $\Lambda$CDM predictions and deepening our understanding of galaxy evolution. 

%%% ACKNOWLEDGMENTS (IF ANY) %%%%%%%%%%%%%%%%%%%%%%%%%%%%%%%%%%%%%%%%

\acknowledgments
We are grateful to the anonymous referee for insightful comments and suggestions that helped improve the manuscript. 
This research was supported by the Korea Astronomy and Space Science Institute under the R\&D program (Project No. 2026-1-831-03) supervised by the Korea AeroSpace Administration.
HK acknowledges the support of the Agencia Nacional de Investigación y Desarrollo (ANID) ALMA grant funded by the Chilean government, ANID-ALMA-31230016.

%%% APPENDICES (IF ANY) %%%%%%%%%%%%%%%%%%%%%%%%%%%%%%%%%%%%%%%%%%%%%

%\appendix
%\section{Appendix Title}

%Some text.

%%% CALL LIST OF REFERENCES (natbib STYLE) %%%%%%%%%%%%%%%%%%%%%%%%%%
\bibliography{jkas-ms}

%\begin{thebibliography}{}

%%% PUT YOUR REFERENCES HERE %%%%%%%%%%%%%%%%%%%%%%%%%%%%%%%%%%%%%%%%

%\bibitem[Salpeter(1955)]{salpeter1955} Salpeter, E. E. 1955, The Luminosity Function and Stellar Evolution, ApJ, 121, 161

%%% END LIST OF REFERENCES %%%%%%%%%%%%%%%%%%%%%%%%%%%%%%%%%%%%%%%%%%

%\end{thebibliography}

\end{document}